\documentclass[trackchanges,twocolumn]{aastex701}

\usepackage{amsmath,amssymb,amsfonts}%
\usepackage{amsthm}%
\usepackage[version=4]{mhchem} % For chemical formulas

\begin{document}

\title{Juno Microwave Radiometer Observations Reveal A Warmer Polar Atmosphere on Jupiter}

\author[0000-0002-6666-5457]{Jiheng Hu}
\affiliation{Department of Climate and Space Sciences and Engineering, University of Michigan, Ann Arbor, MI, USA.}
\email{jihenghu@umich.edu}  

\author[0000-0002-8280-3119]{Cheng Li} 
\affiliation{Department of Climate and Space Sciences and Engineering, University of Michigan, Ann Arbor, MI, USA.}
\email[show]{chengcli@umich.edu}

\author[0000-0002-1972-1815]{Sushil K. Atreya}
\affiliation{Department of Climate and Space Sciences and Engineering, University of Michigan, Ann Arbor, MI, USA.}
\email{atreya@umich.edu}

\author[0000-0001-5834-9588]{Leigh N. Fletcher}
\affiliation{School of Physics and Astronomy, University of Leicester, Leicester, UK.}
\email{leigh.fletcher@leicester.ac.uk}

\author[0000-0002-5440-8779]{Eli Galanti}
\affiliation{Department of Earth and Planetary Sciences, Weizmann Institute of Science, Rehovot, Israel.}
\email{eli.galanti@weizmann.ac.il}

\author[0000-0002-7188-8428]{Tristan Guillot}
\affiliation{Université Côte d’Azur, OCA, Lagrange CNRS, Nice, France.}
\email{tristan.guillot@oca.eu}

\author[0000-0003-4089-0020]{Yohai Kaspi}
\affiliation{Department of Earth and Planetary Sciences, Weizmann Institute of Science, Rehovot, Israel.}
\email{yohai.kaspi@weizmann.ac.il}

\author[0000-0002-5257-9849]{Liming Li}
\affiliation{Department of Physics, University of Houston, Houston, TX, USA.}
\email{lli7@central.uh.edu}

\author[0000-0003-1776-291X]{Yuan Lian}
\affiliation{Aeolis Research, Chandler, AZ, USA.}
\email{lian@aeolisresearch.com}

\author[0000-0002-4552-4292]{Alessandro Mura}
\affiliation{INAF-Istituto di Astrofisica e Planetologia Spaziali, Roma, Italy.}
\email{alessandro.mura@inaf.it}

\author[0000-0001-7871-2823]{Glenn S. Orton}
\affiliation{Jet Propulsion Laboratory, California Institute of Technology, Pasadena, CA, USA.}
\email{glenn.s.orton@jpl.nasa.gov}

\author[0000-0002-8862-8737]{Fabiano A. Oyafuso}
\affiliation{Jet Propulsion Laboratory, California Institute of Technology, Pasadena, CA, USA.}
\email{fabiano@jpl.nasa.gov}

\author[0009-0008-5618-9873]{Maria Smirnova}
\affiliation{Department of Earth and Planetary Sciences, Weizmann Institute of Science, Rehovot, Israel.}
\email{maria.smirnova@weizmann.ac.il}

\author[0000-0002-1978-1025]{J. Hunter Waite}
\affiliation{Department of Physics and Astronomy, The University of Alabama, Tuscaloosa, AL, USA.}
\email{jhwaite@ua.edu}

\author[0000-0003-2804-5086]{Michael H. Wong}
\affiliation{Space Sciences Laboratory, University of California, Berkeley, CA, USA.}
\affiliation{Carl Sagan Center for Science, SETI Institute, Mountain View, CA, USA.}
\email{mikewong@ssl.berkeley.edu} %0000-0003-2804-5086

\author[0000-0002-1558-4948]{Zhimeng Zhang}
\affiliation{Department of Geological and Planetary Sciences, California Institute of Technology, Pasadena, CA, USA.}
\email{zhimeng@caltech.edu}

\author[0000-0003-2242-5459]{Steven M. Levin}
\affiliation{Jet Propulsion Laboratory, California Institute of Technology, Pasadena, CA, USA.}
\email{steven.m.levin@jpl.nasa.gov}

\author[0000-0002-9115-0789]{Scott J. Bolton}
\affiliation{Southwest Research Institute, San Antonio, TX, USA.}
\email{scott.bolton@swri.org}

\correspondingauthor{Cheng Li} 

\begin{abstract}
The intriguing circumpolar cyclone pattern at Jupiter’s poles raises fundamental questions about how these systems are organized vertically and, further, how the planet’s internal heat shapes and sustains them in the absence of solar insolation.
We report recent close-in observations of Jupiter’s north pole acquired by NASA’s Juno Microwave Radiometer (MWR), which achieved comprehensive microwave mapping of the region at an unprecedentedly high resolution. 
Using six-channel measurements from eleven perijove passes (PJ51–PJ61) poleward of $75^\circ$N, we derive polar-mean nadir brightness temperatures and limb-darkening spectra that together point to two equally plausible atmospheric scenarios: (1) a dry-adiabatic profile with slightly depleted ammonia gas at a few bars, or (2) a moist-adiabatic profile with uniform ammonia. 
Markov chain Monte Carlo retrievals yield a deep ammonia abundance of $354.8^{+12.0}_{-11.0}$ ppmv ($\sim3\pm0.1\times$solar) and a water abundance of $1.8^{+1.5}_{-1.1}\times10^{3}$ ppmv ($\sim2.1^{+1.8}_{-1.3}\times$solar), resembling previous estimates at lower latitudes. 
Remarkably, the north pole is found to be 6–7 K warmer than the equator at the 1-bar level, although the inferred difference is close to the 1-$\sigma$ uncertainty level. If confirmed, this result would suggest an enhanced internal heat flux toward the poles, which is consistent with the more intense lightning activity observed at high latitudes. 
\end{abstract}


\section{Introduction}\label{sec:intro} 
In Jupiter’s low to mid-latitudes, rapid rotation organizes the winds into a banded structure known as belts and zones, named for their visual appearance and vorticity \citep{smithJupiterSystemEyes1979}. Toward the poles, zonally oriented bands give way to turbulent weathers and ultimately to polygonal arrangements of circumpolar cyclones \citep{espositoCassiniImagingJupiter2003, boltonJupitersInteriorDeep2017, adrianiClustersCyclonesEncircling2018}. The underlying cause of this transition depends on how Jupiter’s deep atmosphere behaves. If air parcels move freely from depth to the upper layers, their trajectories tend to align with the rotation axis, imprinting angular-momentum structures onto the visible atmosphere \citep{heimpelSimulationEquatorialHighlatitude2005, kaspiDeepWindStructure2009, kaspiObservationalEvidenceCylindrically2023}. Alternatively, if the atmosphere is vertically stratified, with motions confined to isentropic surfaces, the observed latitudinal contrasts arise from interactions between planetary waves and zonal flows \citep{choMorphogenesisBandsZonal1996}.

Investigating the atmospheric thermal and compositional structure across the planet is therefore essential for diagnosing the depth of Jupiter’s weather layer. 
Flyby and ground-based microwave observations have indicated that Jupiter’s equatorial region possesses a vertically well-mixed ammonia vapor distribution with a slight enrichment toward higher altitudes, possibly accompanied by a supper-adiabaticity.\citep{liDistributionAmmoniaJupiter2017, moeckelAmmoniaAbundanceDerived2023, liSuperadiabaticTemperatureGradient2024}. 
In contrast, mid-latitudes display strong vertical ammonia gradients, suggestive of stratification \citep{liDistributionAmmoniaJupiter2017}. 
Jupiter's polar regions, however, remain the least characterized; Earth-based views are highly oblique, as were the views from earlier missions with low-inclination orbits, including the Galileo, Voyager, and Cassini spacecraft, which returned only low-resolution images of haze-covered poles, without revealing any distinct meteorological systems \citep{porcoCassiniImagingJupiters2003, espositoCassiniImagingJupiter2003, maukTransientAuroraJupiter2002}. Prior to the Juno mission, only Pioneer 11 once obtained a polar vantage point but delivered only a snapshot of the polar dynamics \citep{oppScientificResultsPioneer1975}. Moreover, none of these missions carried the multi-wavelength instruments required to resolve the vertical structure.

NASA’s Juno spacecraft, in a high-inclination orbit, now provides continuous pole-to-pole coverage with unprecedented high resolution \citep{boltonJupitersInteriorDeep2017}. Measurements from JunoCam \citep{hansen_junocam_2017} and JIRAM (Jupiter InfraRed Auroral Mapper) \citep{adriani_JIRAM_JovianInfrared_2017} first revealed the mottled morphology of elongated turbulent structures and stable polygonal cyclone clusters \citep{ortonFirstCloseupImages2017, adrianiClustersCyclonesEncircling2018}. However, understanding the evolution \citep{siegelmanPolarVortexCrystals2022, siegelmanMoistConvectionDrives2022, ingersollVorticityDivergenceScales2022}, motion \citep{muraOscillationsStabilityJupiter2021, gavrielOscillatoryMotionJupiters2022}, and stability \citep{liModelingStabilityPolygonal2020,gavrielNumberLocationJupiters2021} of these cyclonic systems has been restricted to depths of only the upper few bars probed in visible and infrared bands, thus, can not determine how deeply these weather systems extend.

The Juno Microwave Radiometer (MWR) uniquely penetrates from millibar levels down to hundreds of bars, measuring both ammonia and temperature profiles \citep{ingersollImplicationsAmmoniaDistribution2017, liWaterAbundanceJupiters2020, boltonMicrowaveObservationsReveal2021, moeckelTempestsTroposphereMapping2025}. As Juno’s periapsis keeps drifting poleward, MWR now achieves its most favorable viewing geometry for polar sounding. Here, we present the mean composition and thermal structure of Jupiter’s northern polar atmosphere, constrained by Juno/MWR’s initial polar measurements, which serve as the foundation for future cyclone-focused studies and provide clues to Jupiter’s global dynamical regime.

\section{Data and Methodology}
\subsection{Juno/MWR Polar Observations}
Approximately 12,000 antenna temperature ($T_A$) measurements were collected by MWR over Jupiter’s north pole (poleward of $75^\circ$N) during eleven Juno perijove passes, spanning PJ51 (May 16, 2023) to PJ61 (May 12, 2024) (Table~\ref{tab:perijovetable}). These measurements were made at favorable low altitudes, 3,000 km to 110,000 km above the visible cloud tops. 
We selected data with at least 85\% of the beam located within an emission angle of 53$^\circ$, where 0$^\circ$ corresponds to the nadir viewing angle.

These multi-orbit $T_A$ measurements were converted to brightness temperatures ($T_b$) through a spectral deconvolution approach \citep{zhangResidualStudyTesting2020}, with the aim of establishing a polar-mean angular dependence model of $T_b$, which is found to be adequately characterized by a quadratic polynomial model \citep{oyafusoAngularDependenceSpatial2020}:
\begin{equation}\label{eq:tb_model}
T_{b,\lambda}(\mu)= [a_\lambda+b_\lambda(1-\mu)+c_\lambda(1-\mu)^2]\cdot\xi_\lambda(\mu),
\end{equation}
\noindent
where $\mu=cos\theta$ ($\theta$ is the emission angle). $a$ (nadir $T_b$), $b$ and $c$ are the polar-mean limb darkening coefficients, depending on the wavelength $\lambda$. $\xi(\mu)$ is the scale function, which equals unity for small emission angles $\mu>0.6$ (53$^\circ$), while applying an additional correction to approach the theoretical modeling when $\mu<0.6$. The limb darkening, $R(\theta)$, is also determined from Eq.~\ref{eq:tb_model}, defined as the percentage reduction in $T_b$ as a function of the emission angle $\theta$:
\begin{equation}
R(\theta) = \frac{T_b(0)-T_b(\theta)}{T_b(0)}\times100\%.
\end{equation}
\noindent

The deconvolution was carried out using an iterative least-squares fitting scheme. At each iteration, the local $T_b$ values computed from the current model (Eq.~\ref{eq:tb_model}) are forward-projected to beam-averaged $T_A$ at each observation point, and the coefficients in Eq.~\ref{eq:tb_model} are subsequently updated based on the residuals between the modeled and observed $T_A$. This process is repeated until the residuals converge. We refer readers to \cite{zhangResidualStudyTesting2020} for detailed descriptions.

Table \ref{tab:tab1} summarizes the deconvolved polar mean nadir $T_b$ and $R(45^\circ)$ for six MWR channels. The nadir $T_b$, from 22 GHz to 0.6 GHz, is 141.1, 192.5, 251.7, 338.8, 477.7, and 908.5 K, respectively, demonstrating that atmospheric temperatures generally increase with depth along a temperature profile close to an adiabat.

\begin{table*}[ht!]
\centering
\caption{\textbf{Nadir brightness temperature and limb darkening of Jovian north pole (75$^\circ$N to 90$^\circ$N).} Deconvolutions were applied to MWR observations during Juno's perijoves PJ51 to PJ61.}
\label{tab:tab1}%
\begin{tabular*}{\textwidth}{@{\extracolsep\fill}cccccc}
\hline
Channel & Frequency & Nadir $T_b$& $R(45^\circ)$ & Sampling Pressure* & Sampling Depth \\
&  (GHz) &  (K) & (\%) &  (bar) & (km) \\
\hline
1 & 0.6   & 908.5$\pm$37.6   & 13.2 & 238.7 & -165 \\
2 & 1.25   & 477.7$\pm$5.2   & 10.0 & 26.9 & -98  \\
3 & 2.6   & 338.8$\pm$1.1   & 6.9 & 8.3 & -63 \\
4 & 5.2   & 251.7$\pm$1.0   & 5.9 & 3.1 & -34 \\
5 & 10  & 192.5$\pm$1.3   & 4.5 & 1.3 & -8 \\
6 & 22  & 141.1$\pm$0.7   & 1.4 & 0.5 & 21 \\
\hline
\end{tabular*}
\raggedright\textsuperscript{*} Sampling pressure is defined as the pressure level where the atmospheric thermal temperature equals the observed brightness temperature, and the depth is the corresponding altitude above the 1 bar pressure level.
\end{table*}

\begin{figure*}[ht!]
    \centering
    \includegraphics[width=0.9\linewidth]{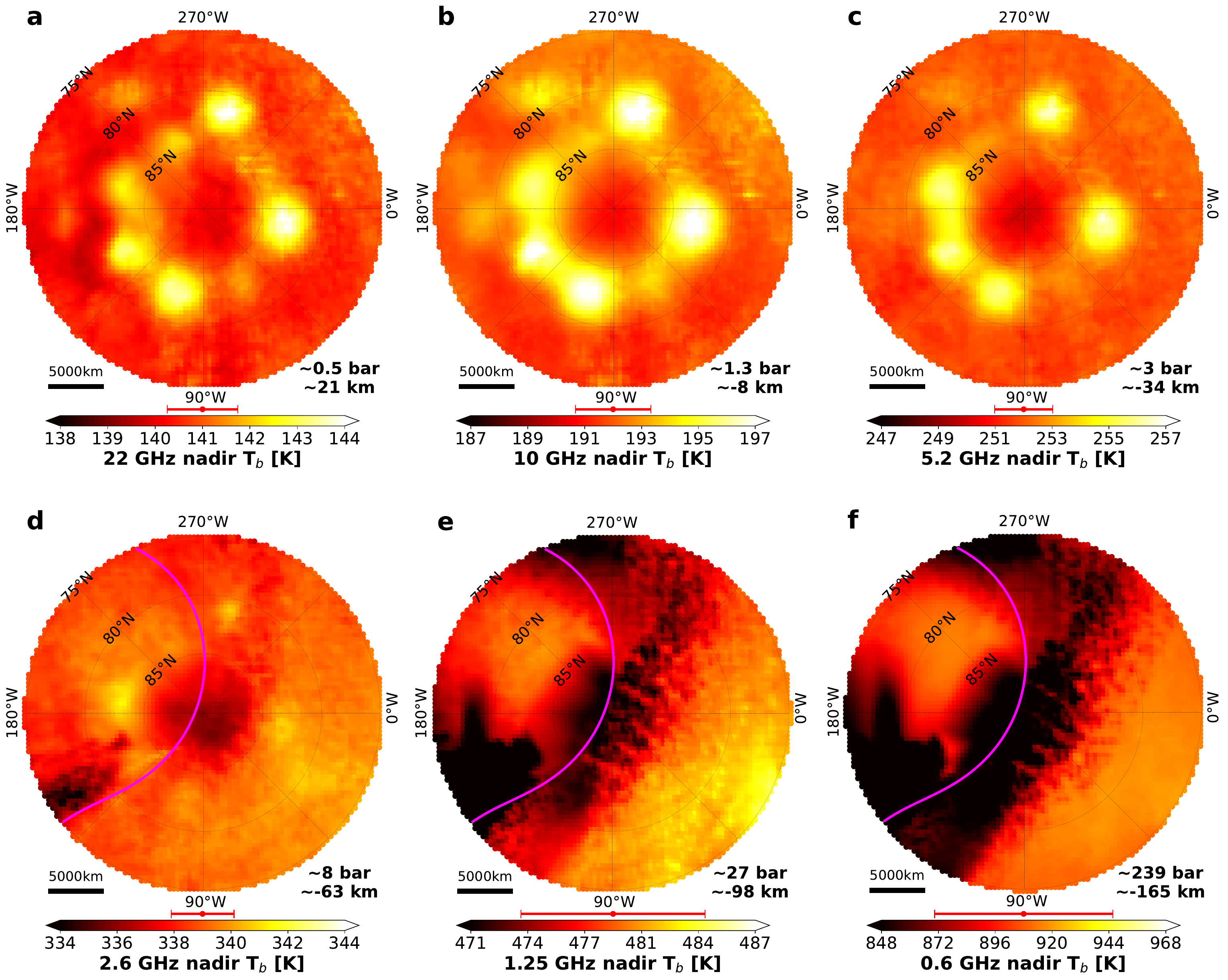}
    \caption{\textbf{Nadir brightness temperatures ($T_b$) in Jupiter's north pole during perijoves PJ51 to PJ61}. Maps at six MWR channels: 22 GHz (\textbf{a}), 10 GHz (\textbf{b}), 5.2 GHz (\textbf{c}), 2.6 GHz (\textbf{d}), 1.25 GHz (\textbf{e}) and 0.6 GHz (\textbf{f}). The magenta line superimposed in panels \textbf{d-f} indicates the main auroral oval, which appears for most perijoves (Fig.~\ref{Ext_aurora_perijove}) and significantly suppresses the brightness temperatures at the three lowest frequencies. Labeled pressure indicates the approximate level where the atmospheric thermal temperature equals the observed brightness temperature; the altitude is the corresponding depth above the 1-bar level. The red error bar denote the area-weighted $T_b$ standard deviation over the domain. The corresponding values for panels \textbf{a} through \textbf{f} are 0.7, 1.3, 1.0, 1.1, 5.2, and 37.6 K, respectively.}
    \label{fig-spatial}
\end{figure*}

Fig.~\ref{fig-spatial} illustrates the nadir $T_b$ maps over the polar region for all six MWR channels, computed as the sum of the polar-mean nadir $T_b$ and the final $T_A$ fitting residuals at each observation point. The error bars shown in Figure 1 represent the polar area-weighted $T_b$ standard deviations of approximately 0.5\%, 0.7\%, 0.4\%, 0.3\%, 1.1\%, and 4.1\% from 22 GHz to 0.6 GHz, respectively. Excluding the aurora-contaminated 0.6-GHz channel, the variability remains below ~1\%, demonstrating that the north polar cap behaves as a spatially uniform region that can be robustly characterized using polar-mean atmospheric properties. This is in contrast to the lower latitudes, where the alternating belt–zone structure produces longitudinally coherent zonal brightness temperature variations of approximately 10–15\% across MWR frequencies \citep{boltonJupitersInteriorDeep2017, liWaterAbundanceJupiters2020}, reflecting distinct dynamical regimes associated with deep zonal jets. 

The Jovian north pole is characterized by a polygonal arrangement of eight circumpolar cyclones (CPCs), which manifest as notable brightness anomalies in microwave frequencies against the polar background.
Five of these CPCs appear significantly brighter than the remaining three, being clearly distinguishable from the polar background even at a depth of approximately 8 bar (Fig.~\ref{fig-spatial}d). 
At the center of the domain lies the north polar cyclone (NPC), which appears as a region of significantly negative brightness temperature anomaly (down to -3 K) relative to the background, suggesting a distinct origin from the surrounding CPCs. Although the thermal anomalies associated with the polar cyclones are physically meaningful, clearly exceeding the instrumental noise floor of 0.5 K, they represent localized regional perturbations superimposed on an otherwise homogeneous background and are therefore not the focus of the present study. We will instead investigate the anomalous cyclonic structures using a differential inversion approach in future work. In addition, studies have shown that the circumpolar cyclones exhibit only small positional displacement ($\sim$400 km) over one year \citep{gavrielOscillatoryMotionJupiters2022}, which is much smaller than the typical cyclone diameter ($\sim$5000 km). Therefore, these small shifts have a negligible influence on the averaged cyclone patterns presented in Figure \ref{fig-spatial}.

Jupiter's auroral region, indicated by the magenta curve in Fig.~\ref{fig-spatial}d-f, exhibits unusually low brightness temperatures due to the enhanced electron density and absorption \citep{hodgesObservationsElectronDensity2020,bhattacharyaHighlyDepletedAlkali2023}, which appear in most of the eleven selected perijoves (Fig.~\ref{Ext_aurora_perijove}). The auroral contamination of atmospheric emission is highest at the 0.6-GHz channel (Fig.~\ref{fig-spatial}e) and decays to minimal levels at the 2.6-GHz channel (Fig.~\ref{fig-spatial}d). 

Although non-atmospheric emissions from lightning, synchrotron, and aurora were flagged and cleaned to the fullest extent possible through a series of data quality assessments \citep{oyafusoAngularDependenceSpatial2020}, complete removal of these interferences is difficult to achieve and access due to the inadequate understanding of the radiative characteristics of the aurora and its variable nature. As a result, we estimate the nadir $T_b$ uncertainty of the 1.25-GHz channel to be 1\% to 2\% and its absolute limb darkening $R(45^\circ)$ uncertainty to be 0.5 to 2.3 K. The lower bound reflects our optimal estimate based on instrument calibration and noise \citep{janssenMicrowaveRemoteSensing2005,janssenMWRMicrowaveRadiometer2017}, while the upper bound represents a conservative precision derived from the observational statistics (Appendix~\ref{appendix_uncertainty}). The 0.6-GHz channel was excluded from our quantitative analysis due to strong auroral contamination and large opacity uncertainties associated with variable alkali metal abundances \citep{bhattacharyaHighlyDepletedAlkali2023, aglyamovAlkaliMetalDepletion2025, Zhang2026-AlkaliMineralDeepVariability}.

In the following manuscript, we report the analysis using a 0.5 K absolute limb darkening uncertainty; then, we discuss the impact if the actual uncertainty is larger due to possible auroral contamination. 

\subsection{Spectral Analysis}

\begin{figure*}[ht!]
    \centering

    \includegraphics[width=\linewidth]{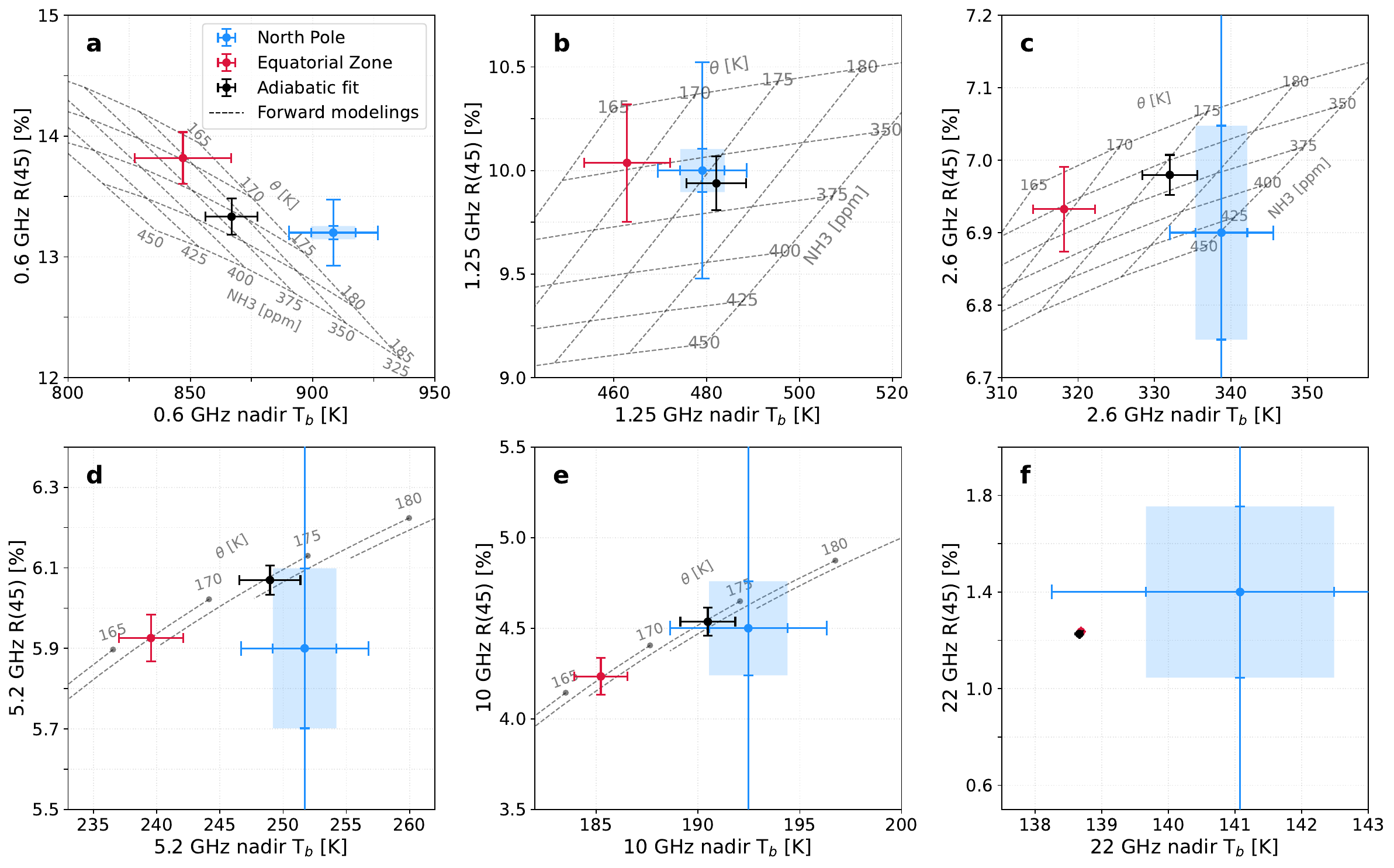}
    \caption{\textbf{Modeled and measured brightness temperatures of Jupiter's north pole (75$^\circ$N$\sim$90$^\circ$N) for all MWR channels.} \textbf{a}, 0.6 GHz, \textbf{b}, 1.25 GHz, \textbf{c}, 2.6 GHz, \textbf{d}, 5.2 GHz, \textbf{e}, 10 GHz, and \textbf{f}, 22 GHz. Dashed lines represent the radiative simulations of nadir $T_b$ and $R(45^\circ)$ as functions of deep (pressure$>$8 bar) NH$_3$ abundance and potential temperature $\theta$ of a dry adiabatic, equilibrium ammonia atmosphere model. Black points denote the best fit of this model, yielding $\theta$= 176.2$\pm1.3$ K and NH$_3$= 362.7$\pm10.2$ ppmv when H$_2$O is set to 2500 ppmv. Blue points represent MWR's polar observations with two uncertainty intervals in nadir $T_b$ (1\% and 2\%) and $R(45^\circ)$ (0.5 K and 2.3 K), corresponding to the optimal and conservative observational precision levels, respectively. Red points denote the reference spectra of Equatorial Zone simulated with the parameters ($\theta$=169 K, NH$_3=351_{-21}^{+22}$ ppmv, H$_2$O=$2.5_{-1.6}^{+2.2}\times10^3$ ppmv) reported in \cite{liWaterAbundanceJupiters2020} but with a gravitational acceleration at the pole. 
    The uncertainties of the EZ and best-fit spectra are quantified through 1000-step Monte Carlo simulations. The simulated variations (grey-dashed) collapse into lines in 5.2 and 10 GHz (\textbf{d–e}) and eventually diminishes in 22 GHz (\textbf{f}), because these frequencies are less sensitive to deep atmospheric parameters due to their limited penetration depths (see Fig. \ref{Ext_weigthing_funcs}b).}
    \label{fig-spectra}
\end{figure*}

Fig.~\ref{fig-spectra} shows a sensitivity analysis to investigate the key parameters that effectively characterize the atmospheric structure at Jupiter’s north pole. We performed a series of radiative transfer simulations to predict the nadir $T_b$ and R(45$^\circ$) across the six MWR channels as functions of the potential temperature and the deep concentration of ammonia. These simulations used an equilibrium condensation model \citep{liMoistAdiabatsMultiple2018} to construct the atmosphere, assuming that the north polar atmosphere follows an ideal dry adiabat and maintains an equilibrium ammonia profile -- meaning that ammonia is vertically well-mixed and depleted solely through condensation to form the ammonia cloud at $\sim$0.7 bar. Nadir $T_b$ and R(45$^\circ$) for the Equatorial Zone (EZ, 0-4$^\circ$N) were also simulated for each channel as a reference using results published early in the Juno mission \citep{liWaterAbundanceJupiters2020}, but with the gravitational acceleration of the polar region. For all the above simulations, we used a fixed equilibrium water vapor profile with a deep abundance set to 2,500 ppmv (parts per million by volume), an estimate at EZ by \cite{liWaterAbundanceJupiters2020}. 

Compared to the EZ reference spectra (red points), observations at the north pole (blue points) exhibit elevated nadir $T_b$ across all channels, suggesting a significant difference over the pole that could be attributed to either a lower ammonia concentration or a warmer air temperature at the layers probed by these channels. The measured R(45$^\circ$) over the north pole is generally comparable to those in EZ, except at 10 GHz, which implies a steeper opacity or temperature gradient at relatively shallow altitudes (pressure$<$8 bar) in the north pole. 
Our simulations show that the 1.25-GHz channel is most sensitive, in a linear fashion, to the deep NH$_3$ concentration and the potential temperature $\theta$. By comparison, the 22-GHz channel exhibits the lowest sensitivity, as it does not penetrate deeper than the ammonia cloud base (Fig. \ref{Ext_weigthing_funcs}b), thus its opacity is predominantly governed by the ammonia relative humidity within the cloud layer.

A large discrepancy is found between our simulation and observation at 0.6 GHz (Fig.~\ref{fig-spectra}a), which is possibly due to a combination of (1) severe radio frequency interference from the polar aurora \citep{bhattacharyaJupitersAuroralIonosphere2025}, (2) synchrotron radiation \citep{levinModelingJupitersSynchrotron2001}, and (3) varying alkali metal abundances \citep{bhattacharyaHighlyDepletedAlkali2023, aglyamovAlkaliMetalDepletion2025}. We display the calculation results but did not attempt to fit the 0.6-GHz channel data throughout this study.

Under the assumptions of a dry adiabat and an equilibrium ammonia atmosphere, we attempted to fit the observations across all MWR channels (excluding 0.6 GHz) by varying the potential temperature and deep ammonia abundance. The best-fit solution (Fig.~\ref{fig-spectra}, black dots) suggests a deep ammonia abundance of 362.7$\pm10.2$ ppmv and a potential temperature of 176.2$\pm1.3$ K. This solution accurately reproduces the measured R(45$^\circ$) across all channels, with absolute errors within 0.5 K. For nadir $T_b$, the model agrees with the observation at 1.25 GHz within a 1\% error; yet, it significantly underestimates the four higher-frequency channels, with biases of 1\% to 2\%.
These discrepancies relative to the nominal model suggest that two alternative atmospheric structures may exist in Jupiter's north pole: (1) the ammonia vapor is slightly depleted at a few bars, or (2) the atmospheric temperature profile is closer to a moist adiabat (warmer than the adiabatic case at a few bars).

\subsection{Atmospheric Retrieval Algorithm}

\subsubsection{Two temperature-ammonia scenarios}\label{two-scenarios}
The observed brightness temperature is affected by temperature and composition. Considering that there are only five MWR frequencies that could be leveraged to constrain the polar atmospheric state, and with the aim of characterizing the profiles in a thermodynamic way, in this study, we only tried two simplest end-member adiabatic models to retrieve the polar mean atmospheric structure. The first model assumes a dry adiabatic profile with a depleted NH$_3$ abundance at a few bars; while the second model uses an equilibrium NH$_3$ profile but with a moist adiabatic temperature, accounting for the release of latent heat due to water condensation at a few bars. Such thermodynamic models, which differ from the stochastically process-based models \citep{liDistributionAmmoniaJupiter2017, moeckelAmmoniaAbundanceDerived2023}, are more physically based and do not strive to create a more delicate structure, but emphasize yielding a more physically interpretable solution. In the moist-adiabat case, we attribute the change in atmospheric lapse rate to latent heating from water condensation, expected near the 4-7 bar level. Although the retrieved water abundance primarily represents conditions near $\sim$5 bar, we use it as a proxy for the deep abundance by assuming a well-mixed deep atmosphere. This treatment follows the Occam’s razor principle: we adopt the minimum-complexity model required to fit the MWR observations unless the observations or the physics indicate otherwise.

The condensing layer of ammonia is at $\sim$0.7 bar; therefore, NH$_3$ condensation alone cannot explain the slight opacity shortage at a few bars. In practice, we model the slight depletion of ammonia gas by applying a linear gradient or a step function of altitude \citep{depaterPeeringJupitersClouds2016, liWaterAbundanceJupiters2020} at a few bars. The rationale for such depletion is threefold: (1) interaction with H$_2$S to form the solid NH$_4$SH cloud at $\sim$2.5 bar (NH$_3$ + H$_2$S $\rightarrow$ NH$_4$SH) \citep{liMoistAdiabatsMultiple2018, depaterPeeringJupitersClouds2016}; (2) dissolution of ammonia vapor in the water cloud at $\sim$7 bar \citep{weidenschillingAtmosphericCloudStructures1973, atreyaComparisonAtmospheresJupiter1999, guillotStormsDepletionAmmonia2020}; (3) transport to deeper layers by dry downdrafts \citep{liSimulatingNonhydrostaticAtmospheres2019, fletcherJupitersTemperateBelt2021}. 

In addition, to adequately fit the 22-GHz channel, following \cite{depaterPeeringJupitersClouds2016}, we introduce an upper limit for the relative humidity of ammonia, RH$_\text{max}$, ranging from 0 (dry) to 1 (fully saturated), to restrict the NH$_3$ concentration at and above the NH$_3$ cloud. The rationale is that the excess 22-GHz brightness temperature ($T_b$) relative to the nominal model indicates an undersaturation of ammonia vapor in the ammonia cloud layer. This may result from ammonia depletion relative to temperature or from a locally warmer upper troposphere with vertically uniform ammonia. The parameter RH$_{\max}$ can effectively constrain any continuum state between these two scenarios without attempting to resolve their degeneracy. On Earth, a fully saturated atmosphere is typically found only near the clouds of the marine boundary layer or in actively convecting regions \citep{xuTropicalAtmosphereConditionally1989, brethertonGravityWavesCompensating1989}. On Jupiter, ammonia is pervasively found to be subsaturated \citep{achterbergCassiniCIRSRetrievals2006, fletcherMidinfraredMappingJupiters2016, grassiCloudsAmmoniaJupiters2021}. 

Therefore, for the dry adiabatic case, we use five parameters to characterize the atmospheric structure: (1) $x\text{NH}_3$, the deep (p $>$ 8 bar) ammonia abundance, in ppmv; (2) T$_\text{1bar}$, the temperature at the 1 bar level; (3) RH$_\text{max}$, an upper limit of ammonia relative humidity; (4) $\Delta\Gamma=\Delta\frac{\text{dln}x\text{NH}_3}{\text{dlnP}}$, a dimensionless lapse rate to exert a linear ammonia gradient (negative value indicates a decrease with altitude); and (5) $p_d$, the pressure level where the depletion starts from the bottom. For the moist adiabatic case, we achieve a good fit across all channels by varying only three parameters: (1) $x\text{H}_2$O, the deep water abundance in ppmv, (2) T$_\text{1bar}$, and (3) RH$_\text{max}$ for ammonia. For both cases, NH$_3$ abundance and potential temperature ($\theta$) in the deep layer should be consistent, as their main deviations occur in a few bars, where the condensation of water and the depletion of ammonia begin to make a difference. 

We first retrieve, under the dry adiabatic assumption, the deep ammonia abundance, with a deep water abundance $x\text{H}_2$O set to 2,500 ppmv, a reference value of the Equatorial Zone \citep{liWaterAbundanceJupiters2020}. Then, we perform the moist adiabatic retrieval to vary the deep water abundance and, accordingly, the moist adiabatic temperature profile. The rationale is that the deep atmospheric properties are primarily constrained by the 1.25-GHz brightness temperatures, whose weighting functions peak in the deep atmosphere and are therefore largely insensitive to the thermodynamic assumptions at shallower levels (Fig~\ref{Ext_weigthing_funcs}b). Figure \ref{fig:ext_sensi_xH2O} illustrates that the 1.25-GHz brightness temperature modeled in the moist adiabat model is largely insensitive to deep water abundance but remains strongly dependent on potential temperature and deep ammonia abundance. Thus, the 1.25-GHz observations primarily constrain the deep ammonia abundance and thermal structure, while the deep water abundance is mainly constrained by higher-frequency channels. Therefore, the deep ammonia abundance retrieved under the dry adiabat assumption can be directly adopted in the moist adiabat case, with only minimal additional uncertainty. 

% A Markov chain Monte Carlo (MCMC) approach \citep{goodmanEnsembleSamplersAffine2010} is adopted to explore the parameter space and determine the posterior probability distributions of the aforementioned variables, aiming to reproduce the observed brightness temperatures at five MWR frequencies (1.25, 2.6, 5.2, 10, and 22 GHz) and across four emission angles ($0^\circ$, $15^\circ$, $30^\circ$, and $45^\circ$). These 20 channel-angle radiance measurements, in conjunction with the corresponding error covariance matrix, provide simultaneous constraints on both the nadir brightness temperature and the limb darkening characteristics over the polar region. The detailed retrieval implementation is documented in Appendix \ref{appendix_MCMC}. The prior distributions of parameters for the dry and moist adiabatic scenarios are presented in Fig.~\ref{Ext1_MCMC_dry_corners} and Fig.~\ref{Ext2_MCMC_moist_corners}, respectively.
\subsubsection{Markov chain Monte Carlo inversion}
We opted to use the Bayesian statistical scheme to explore the parameter space under two atmospheric thermodynamic assumptions. The state vectors allowed to vary, as introduced above, are
\begin{equation}\label{mcmc-parameters}
\boldsymbol{\Theta} = 
\begin{cases} 
    [x\text{NH}_3,\,\text{T}_\text{1bar},\,\text{RH}_{max}, \,\Delta\Gamma, \,p_d] & \text{dry adiabat} \\
    [x\text{H}_2\text{O},\,\text{T}_\text{1bar},\,\text{RH}_{max}] & \text{moist adiabat}
\end{cases}
\end{equation}

The scheme aims to explore the posterior distributions of these parameters, by which the determined atmospheric structure fits the observational spectra of twenty channel angles derived from Eq.\ref{eq:tb_model}:
\begin{equation}\label{y-parameters}
\Tilde{\mathbf{Y}}=[T_b^{1.25, 0^\circ}, \,T_b^{1.25, 15^\circ}, ... , \, T_b^{\lambda_k, \theta_k} ...\,,\, T_b^{22, 45^\circ}] \\.
\end{equation}
\noindent
Here, the superscripts denote the channel frequency $\lambda_k\in \{1.25, 2.6, 5.2, 10.0, 22.0\}$ (GHz) and the emission angle $\theta_k\in \{0^\circ, 15^\circ, 30^\circ, 45^\circ\}$. These 20 channel-angle $T_b$ measurements provide simultaneous constraints on both the nadir brightness temperature and the limb darkening characteristics over the polar region.

An equilibrium cloud condensation model \citep{liSimulatingNonhydrostaticAtmospheres2019} was applied to construct atmospheric profiles of temperature, water, and ammonia according to each state of $\boldsymbol{\Theta}$ under the corresponding thermodynamic assumptions (dry/moist adiabat). The temperature profiles were constructed from bottom to top by integrating the temperature lapse rate layer-by-layer, which is determined by the adiabatic theory of multiple condensing species \citep{liMoistAdiabatsMultiple2018}. Taking into account the opacity of NH$_3$, H$_2$O, and the H$_2$/He collision-induced absorptions, radiative transfer modeling is performed for each $\boldsymbol{\Theta}$ state to obtain the $T_b$ spectra, $\mathbf{Y}$.

% According to the Bayesian statistical theorem, the posterior is determined by the likelihood of observing $\mathbf{Y}$ given $\boldsymbol{\Theta}$, $P(\mathbf{Y}|\boldsymbol{\Theta})$, the prior probability of the parameters $P(\boldsymbol{\Theta})$, and the prior probability of a measurement $\mathbf{Y}$, $P(\mathbf{Y})$.

% \begin{equation}\label{bayesian}
% \text{ln}P(\boldsymbol{\Theta}|\Tilde{\mathbf{Y}})= \text{ln}P(\mathbf{Y}|\boldsymbol{\Theta}) + \text{ln}P(\boldsymbol{\Theta}) - lnP(\mathbf{Y})\\.
% \end{equation}
% The application of logarithmic form allows us to compute the logarithmic probability for each channel and parameter in a linear summary form.

Under the Gaussian distribution assumption, the likelihood follows the relationship:
\begin{equation}
\text{ln}P(\mathbf{Y}|\boldsymbol{\Theta}) \propto -\frac{1}{2}[(\mathbf{Y}-\mathbf{\Tilde{Y}})^\text{T}\boldsymbol{\Sigma}
^{-1}(\mathbf{Y}-\mathbf{\Tilde{Y}})]\\.
\end{equation}
\noindent
Here, the error covariance matrix $\boldsymbol{\Sigma}$ is applied to penalize the departure of any state of $\mathbf{Y}$ with respect to the truth $\Tilde{\mathbf{Y}}$. %, which imposes constraints on errors in both the absolute values and limb darkening.
$\boldsymbol{\Sigma}$ is a block diagonal matrix composed of six sub-blocks, each of which is a full matrix:% as illustrated in Fig.~\ref{Ext4_error_covariance_matrix}, 
\begin{equation}\label{covariance}
\Sigma_{i,j}=T_b^{\lambda(i)}T_b^{\lambda(j)}\text{var}(\Delta)\delta_{\lambda(i)\lambda(j)}+\text{var}(\epsilon)\delta_{ij}\\.
\end{equation}
\noindent
Where, $i$ is the index of channel angles. $\lambda(i)$ denotes the corresponding channel frequency. $\delta_{ij}$ is the Kronecker delta. The variances of systematic error, $\text{var}(\Delta)$, and random measurement noise, $\text{var}(\epsilon)$, in the observations are set to 1\% and 0.5 K, respectively, representing an optimistic precision level of the MWR instrument.

% $P(\boldsymbol{\Theta})$ is the prior knowledge of the parameters we have before inference, which could be empirical theory or in situ measurements. 
In this study, priors were set referring to estimates in the EZ \citep{liWaterAbundanceJupiters2020}, with $300_{-100}^{+100}$ ppmv for the deep ammonia abundance, 2,500 ppmv for the deep water abundance ($\sigma=10,000$ ppmv, valid range 0--10,000 ppmv), and $169_{+10}^{-10}$ K for $T_\text{1bar}$. All of them are assumed to follow a Gaussian distribution with large standard deviations, such that they apply relatively loose constraints compared to MWR measurements. For other parameters, the priors are shown in Figs.~\ref{Ext1_MCMC_dry_corners} and \ref{Ext2_MCMC_moist_corners}. 

We adopt an affine-invariant Markov chain Monte Carlo (MCMC) ensemble sampler, the emcee \citep{foreman-mackeyEmceeMCMCHammer2013}, to sample the parameter space. 
We used twelve walkers to run the MCMC in parallel and found that good convergence can be reached within 10,000 steps. We excluded the first 3,000 states of the chain as the burn-in stage and inferred the posterior distribution of parameters from the remaining 7,000 states. % Figs.~\ref{Ext1_MCMC_dry_corners} and \ref{Ext2_MCMC_moist_corners} show the joint and marginal probability distributions of the parameters as determined by the MCMC inversion over the north pole. Fig.~\ref{Ext_weigthing_funcs} presents the ammonia profiles and the corresponding atmospheric weighting functions as retrieved under both adiabatic assumptions. Fig.~\ref{Ext3_LUT_dry_adb} shows the fit performance of solutions under different atmospheric structure assumptions. 

\section{Results}

\begin{figure*}[ht!]
    \centering
    \includegraphics[width=0.8\textwidth]{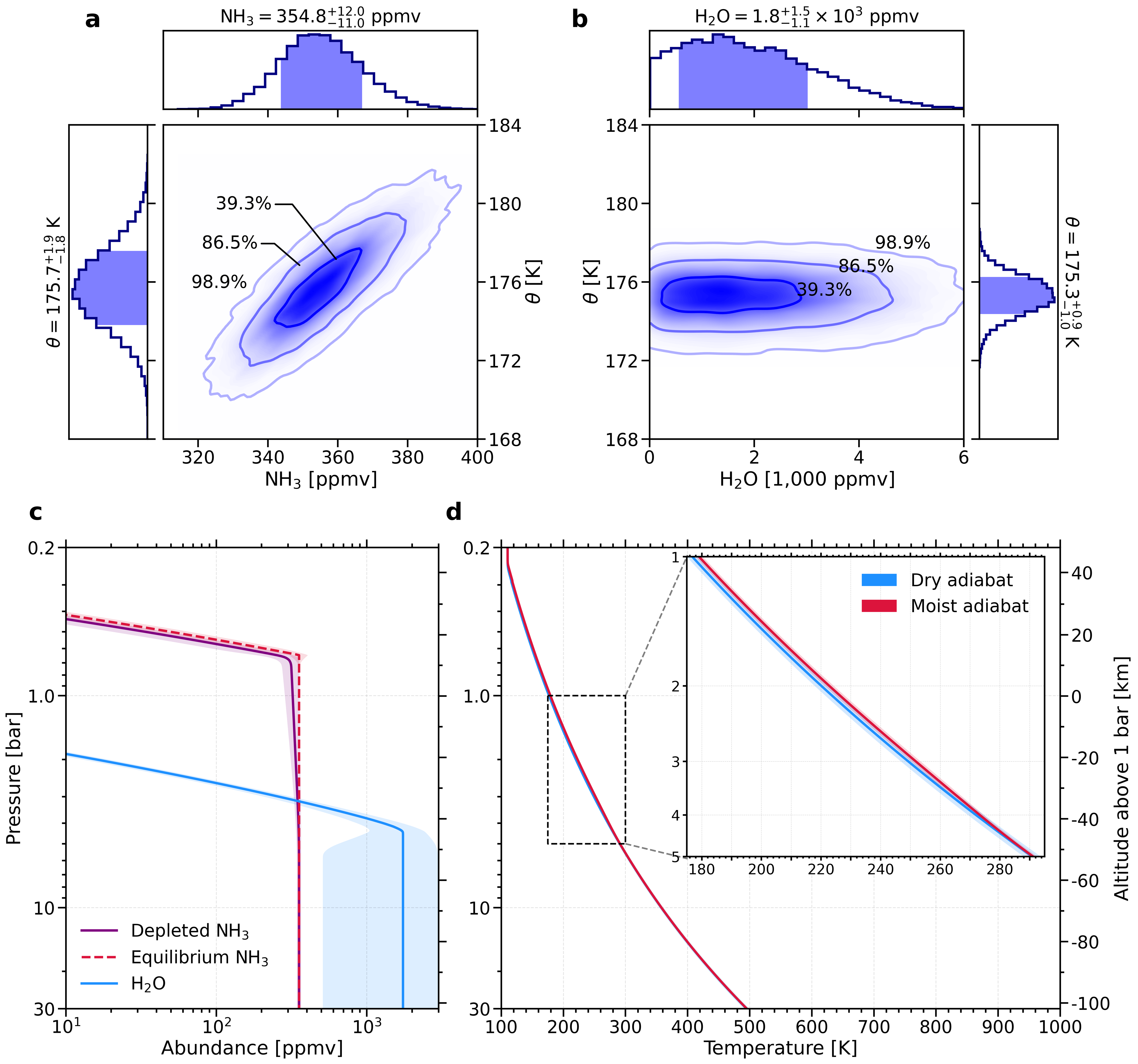}
    \caption{ \textbf{Retrieved profiles of the atmospheric temperature and ammonia and water abundances.}
    \textbf{a,} statistical constraints on deep ammonia abundance and potential temperature ($\theta$) under the dry adiabatic assumption.
    \textbf{b,} statistical constraints on deep water abundance and deep $\theta$ under the moist adiabatic assumption.
    Blue contours in the joint probability distributions denote the 1$\sigma$ (39.3\%), 2$\sigma$ (86.5\%), and 3$\sigma$ (98.9\%) confidence intervals.
    Marginal probability distributions are shown as histograms, with median values and corresponding 68\% confidence intervals indicated by blue shading.
    \textbf{c,} vertical profiles of water vapor (blue solid), equilibrium ammonia (red dashed), and depleted ammonia (purple solid) concentrations.
    \textbf{d,} temperature profiles based on dry (blue) and moist (red) adiabatic thermodynamic models.
    The inset in \textbf{d} expands the 1–5 bar region to highlight differences between the two profiles.
    Shaded in panels \textbf{c–d} are the 1$\sigma$ uncertainty envelopes, derived from the last 7000 steps of the Markov chain.
    }
    \label{fig-jointpdf-profiles}
\end{figure*}

Under the dry adiabatic assumption, the posterior constraint on $x\mathrm{NH_3}$, after marginalizing over the other free parameters, is $354.8^{+12.0}_{-11.0}$ ppmv, which is approximately $3\pm0.1$ times the solar N/H ratio \citep{asplundChemicalCompositionSun2009}. The deep layer potential temperature is constrained to be 175.7$_{-1.8}^{+1.9}$ K. In this analysis, the superscript and subscript represent the 68\% (1$\sigma$ uncertainty) confidence interval centered on the median value. The posterior distribution of the potential temperature $\theta$ (equal to T$_\text{1bar}$ in the dry adiabatic case) and the deep ammonia abundance $x\text{NH}_3$ are reported in Fig.~\ref{fig-jointpdf-profiles}a. The significant positive correlation between $\theta$ and the NH$_3$ concentration indicates a strong degeneracy, suggesting that the observed radiance could be explained by either an increase in temperature or a decrease in atmospheric opacity. The medium ammonia profile sampled by the MCMC is presented as a solid purple curve in Fig.~\ref{fig-jointpdf-profiles}c. The corresponding temperature profile is shown as the solid blue curve in Fig.~\ref{fig-jointpdf-profiles}d.

The RH$_{max}$ is determined to be $0.71_{-0.15}^{+0.16}$ with respect to a saturated value of unity, as a necessary measure to suppress the NH$_3$ opacity at altitudes above the ammonia cloud. In addition, a negative gradient of $\Delta\Gamma=-0.06\pm0.06$ is necessarily applied to the altitude of the $p_d=4.9\pm1.0$ bar, which is responsible for the evident depletion of NH$_3$ concentration from several bars to the visible cloud top (Fig.\ref{fig-jointpdf-profiles}c, solid purple line). Such depletion leads to downward shifts in atmospheric weighting functions at 5.2 and 10 GHz relative to the equilibrium profile (Fig.~\ref{Ext_weigthing_funcs}), meaning that they probe deeper than in the nominal model. 

The posterior distributions of $x\text{NH}_3$, T$_\text{1bar}$, RH$_{max}$ and $\Delta\Gamma$ are remarkably different from their prior distributions (Fig.~\ref{Ext1_MCMC_dry_corners}), indicating significant observational constraints compared to priors and that our understanding of these parameters based on previous findings in the Equatorial Zone \citep{liDistributionAmmoniaJupiter2017, liWaterAbundanceJupiters2020} is largely biased. In contrast, the posterior of $p_d$ is similar to its prior, which implies that no additional information is contained in the observation for this parameter.
The Monte Carlo sampling analysis in Fig.~\ref{Ext3_LUT_dry_adb} demonstrates that, compared to the vertically uniform-mixed ammonia vapor profile, a depleting gradient occurring over a few bars significantly improves the fit to $T_b$ observations at 5.2 and 10 GHz. Meanwhile, by introducing the parameter of RH$_{max}$, we substantially improve the fit to the 22-GHz $T_b$.

For the moist adiabat case, we fix the deep ammonia abundance obtained in the dry adiabat case and vary the $x\text{H}_2$O, T$_\text{1bar}$, and RH$_{max}$. The vertical ammonia profile is therefore regulated by the moist adiabatic temperature profile to maintain thermodynamic equilibrium. The deep water abundance inferred by MCMC is $x\text{H}_2$O$=1.8_{-1.1}^{+1.5}\times10^3$ ppmv or $2.1_{-1.3}^{+1.8}$ times the solar O/H ratio, with a long tail towards large values (Fig.~\ref{fig-jointpdf-profiles}b). We obtained a T$_\text{1bar}$ of $177.9^{+1.4}_{-1.1}$ K, corresponding to a deep-layer (p $>$ 8 bar) potential temperature of $175.3^{+0.9}_{-1.0}$ K, which is consistent with the value in the dry adiabat case. In the moist adiabat case, the temperature profile is about 2 to 3 K higher than in the dry adiabat case at pressure levels between 0.2 and 4 bars (inset of Fig.~\ref{fig-jointpdf-profiles}d).

We further verify using a sensitivity test shown in Figure~\ref{sensitive_moist_NH3} that retrieving deep ammonia abundance together with all other parameters yields nearly identical parameter estimates to those obtained when the deep ammonia abundance is fixed to the value inferred from the dry-adiabatic case. The retrieved deep ammonia abundance is $356.0_{-10.1}^{+10.8}$ ppmv compared to $354.8_{-11.0}^{+12.0}$ ppmv for the dry-adiabatic case, and the deep potential temperature is $176.1_{-1.5}^{+1.6}$ K compared to $175.7_{-1.8}^{+1.9}$ K for the dry-adiabatic case. This exercise demonstrates the claim made in Section \ref{two-scenarios} that MWR 1.25-GHz channel observations dictate the deep atmospheric properties, and our treatment of fixing ammonia abundance for the moist adiabatic case induced minimal differences.

Physically interpreting, the latent heat released from such an amount of water is responsible for the excessive brightness observed at 5.2 and 10 GHz (Fig.~\ref{fig-spectra}) relative to the dry adiabatic fit. The relatively smaller uncertainty of $\theta$ in the moist adiabatic case is due to only three parameters being varied in the retrieval, as we fixed the deep ammonia abundance derived from the dry adiabatic case, thereby reducing the number of free parameters. We also obtained a tight constraint on the ammonia abundance at altitudes above the cloud top with RH$_{max}=0.71\pm0.15$ (Fig.~\ref{Ext2_MCMC_moist_corners}), which is consistent with that of the dry adiabatic case. Monte Carlo sampling analysis also shows that, by applying the moist adiabatic profile, the observations at 5.2 and 10 GHz are well reproduced with a bias within the 1\% error of $T_b$ and the 0.5 K error of limb darkening (Fig.~\ref{Ext3_LUT_dry_adb}).

\section{Discussion and Conclusion}\label{sec12}

\subsection{Deep ammonia abundance} 
\begin{figure*}[htbp]
    \centering
    \includegraphics[width=1\linewidth]{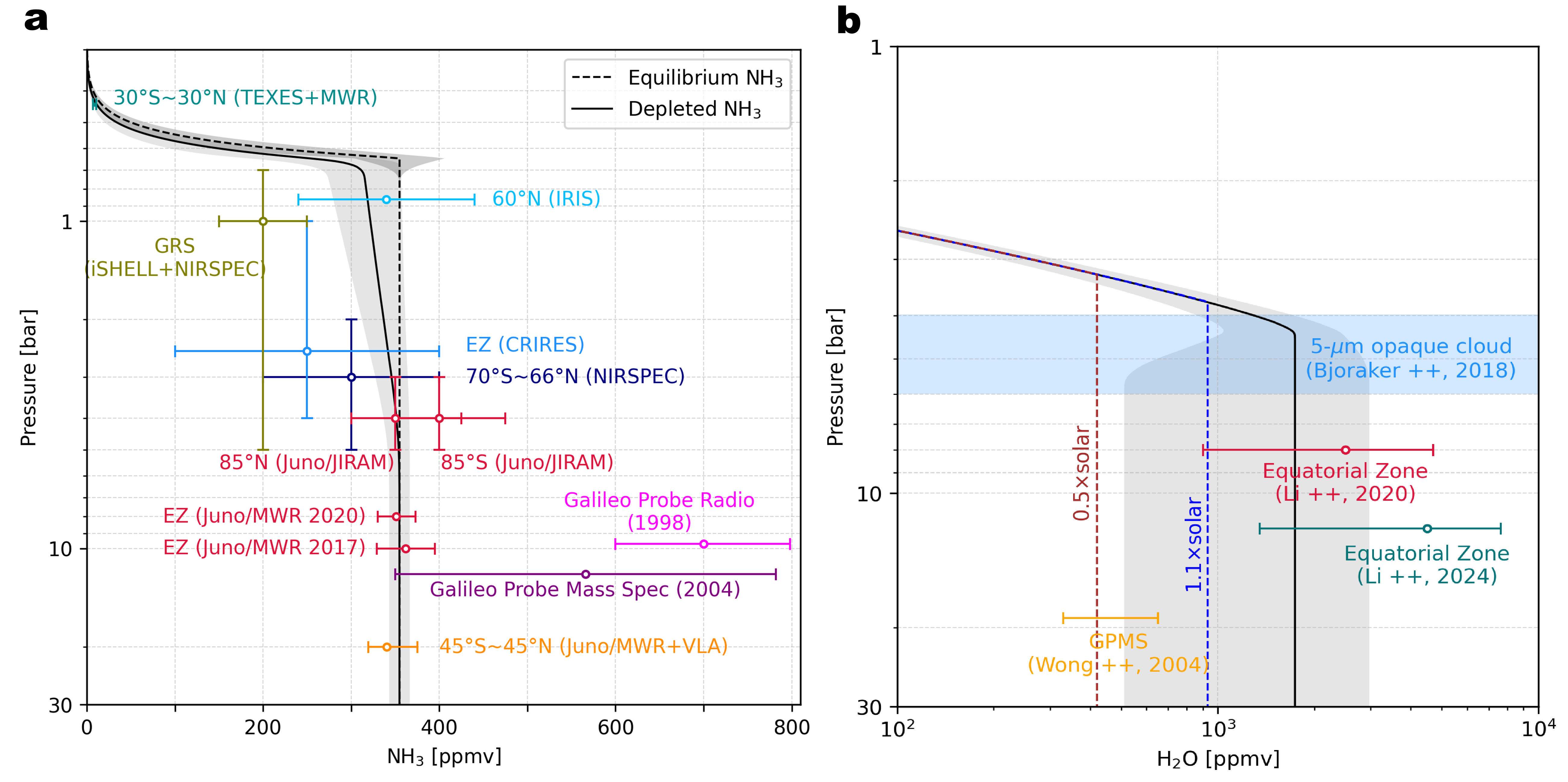}
    \caption{\textbf{Cross-comparison of ammonia and water abundance estimates on Jupiter.} \textbf{a,} Ammonia profile of different regions of Jupiter from varies studies, the whiskers represent the uncertainty or range of the measurements. The solid and dashed profiles denote the polar ammonia profiles of two different adiabatic cases, corresponding to those shown in Fig.~\ref{fig-jointpdf-profiles}c.
    The instruments quoted here are Juno/MWR in the Equatorial Zone (EZ) \citep{liDistributionAmmoniaJupiter2017, liWaterAbundanceJupiters2020}; Galileo Probe Mass Spectrometer (GPMS) \citep{wongUpdatedGalileoProbe2004} and Galileo Probe radio attenuation signal \citep{folknerAmmoniaAbundanceJupiters1998} at the 5-$\mu$m 'hot spot'; Juno/MWR and Very Large Array (VLA) \citep{moeckelAmmoniaAbundanceDerived2023}; The Cryogenic High-resolution Infrared Spectrograph (CRIRES) at the Very Large Telescope (VLT) \citep{gilesAmmoniaJupitersTroposphere2017}; Gemini/TEXES and Juno/MWR \citep{fletcherJupitersEquatorialPlumes2020}; SOFIA/FORCAST and Voyager/IRIS 17–37 $\mu$m spectroscopy \citep{fletcherJupitersParaH2Distribution2017}; 5-$\mu$m spectra of IRTF/iSHELL and Keck II/NIRSPEC at the Great Red Spot (GRS) \citep{bjorakerGasCompositionDeep2018}; Keck II/NIRSPEC for 66$^\circ$N $\sim$ 70$^\circ$S \citep{bjorakerSpatialVariationWater2022}; Juno JIRAM data for a global measurement \citep{grassiSpatialDistributionMinor2020}.
    \textbf{b,} Water abundance estimated by various studies, black solid line indicates the water vapor profile under moist adiabatic case, with grey shaded region denoting its 1$\sigma$ envelope. Dashed lines denote the lower limits of 0.5$\times$ solar \citep{wongDeepCloudsJupiter2023} and 1.1$\times$ solar \citep{bjorakerGasCompositionDeep2018} derived from 5-$\mu$m spectroscopic data, respectively; Red and turquoise solid error bars represent the previous Juno/MWR estimates at the equator in \cite{liWaterAbundanceJupiters2020} and \cite{liSuperadiabaticTemperatureGradient2024}, respectively;  Orange error bar represents the entry measurements of GPMS into the dry 5-$\mu$m hot spot \citep{wongUpdatedGalileoProbe2004}; The blue shaded region represents the deep cloud layer responsible for the opaque layer near 5 bar inferred from 5-$\mu$m spectroscopy of the Great Red Spot \citep{bjorakerGasCompositionDeep2018}.
    }
    \label{fig:nh3_comparison}
\end{figure*}

Fig.~\ref{fig:nh3_comparison}a provides a comparative summary of ammonia abundance at different altitudes and locations on Jupiter from several featured studies. Our estimate of the deep ammonia abundance (pressure$>$5 bar) at the Jovian north pole is $354.8_{-11.0}^{+12.0}$ ppmv, in agreement with the range of values in the equatorial zone, $362\pm33$ ppmv, an initial estimate from MWR's first orbit measurement \citep{liDistributionAmmoniaJupiter2017}. The estimate also coincides with the updated value of $351_{-21}^{+22}$ ppmv in the equatorial zone \citep{liWaterAbundanceJupiters2020} and the range of $340.5_{-21.2}^{+34.8}$ ppmv retrieved between $45^\circ$S and $45^\circ$N \citep{moeckelAmmoniaAbundanceDerived2023}. Compared with studies prior to Juno, our result is close to the lower bound of the measurement from the Galileo Probe Mass Spectrometer ($566\pm216$ ppmv) \citep{wongUpdatedGalileoProbe2004} and is nearly two times smaller than the estimate of the Galileo Probe radio attenuation ($700\pm100$ ppmv) \citep{folknerAmmoniaAbundanceJupiters1998}. 

At higher latitudes, our ammonia vapor profiles are consistent with estimates of $\sim 350_{-50}^{+75}$ ppmv around 85$^\circ$N and $\sim 400_{-50}^{+75}$ ppmv around 85$^\circ$S, which are inferred from the Juno/JIRAM 5-$\mu$m measurements sensing 3 to 5 bars \citep{grassiSpatialDistributionMinor2020}. Our polar estimates also show good agreement with the measurements of the Voyager/IRIS 10-$\mu$m spectra, 340$\pm100$ ppmv at $\sim$0.86 bar for latitude around 60$^\circ$N \citep{fletcherJupitersParaH2Distribution2017}.

We also make comparisons with the retrievals from ground-based telescopes. The 5-$\mu$m measurements of Keck II/NIRSPEC suggested a general ammonia concentration range of 200 to 400 ppmv between 66$^\circ$N and 70$^\circ$S \citep{bjorakerSpatialVariationWater2022}; retrievals of 5-$\mu$m spectrum from the cryogenic high-resolution infrared spectrograph (CRIRES) manifested a general range of 100 to 400 ppmv in the EZ and SEB regions at 1 to 4 bars \citep{gilesAmmoniaJupitersTroposphere2017}; a combined use of 5-$\mu$m measurements from IRTF/iSHELL and the Keck II/NIRSPEC reported a 200 $\pm$ 50 ppmv ammonia between 0.7 and 5 bars in the Great Red Spot (GRS) \citep{bjorakerGasCompositionDeep2018}. These studies are consistent with the findings of \cite{grassiSpatialDistributionMinor2020}, which reported a relatively constant ammonia concentration of approximately 300 ppmv between 25$^\circ$N and 70$^\circ$N, while the ammonia concentration is not very constant in the southern hemisphere, being higher closer to the south pole. Our estimates at these layers (p $<$ 5 bar) are within the upper limit of these lower-latitude estimates, which, to some extent, consolidate the increasing trend in ammonia abundance at higher altitudes (beyond 70$^\circ$ N) toward the north pole, as reported by various observations \citep{grassiSpatialDistributionMinor2020, fletcherJupitersParaH2Distribution2017}. 

Above the ammonia ice cloud top, the polar estimate is comparable to the value of $\sim8.8\pm2$ ppmv at $\sim$0.44 bar determined between 30$^\circ$S and 30$^\circ$N \citep{fletcherJupitersEquatorialPlumes2020}, which indicates that, above the cloud top, the ammonia abundance does not vary much from the equator to the pole. Although these measurements vary in location and depth, the comparisons show comparable and reasonable results for our estimation and illustrate the spatial variability in the global atmospheric ammonia distribution.

\subsection{A depleted polar ammonia profile?}

The presence of a depleted NH$_3$ profile is commonly found on Jupiter (except for the equator) and is used as an alternative model with respect to a fully equilibrium model to fit the spectral data from VLA \citep{depaterPeeringJupitersClouds2016, moeckelAmmoniaAbundanceDerived2023}, CRIRES \citep{gilesAmmoniaJupitersTroposphere2017}, and IRTF/TEXES \citep{blainMappingJupitersTropospheric2018}. 
The preliminary explorations \citep{ingersollImplicationsAmmoniaDistribution2017,boltonJupitersInteriorDeep2017} based on the MWR illustrated a global zonal averaged ammonia vertical distribution between 40$^\circ$S and 40$^\circ$N, which revealed that uniform-mixed ammonia is confined to a region below the 60-bar level, except for the uniform equatorial plume. Previously, such a well-mixed domain was too deep to be probed by the VLA. However, Juno/MWR observations have extended access to this region over a broader latitude range (60$^\circ$S--60$^\circ$N) \citep{boltonJupitersInteriorDeep2017, zhangResidualStudyTesting2020, moeckelAmmoniaAbundanceDerived2023}.

Our retrieval allows for a solution in which ammonia in the north polar region is depleted above the water cloud level, with a transition pressure of 4.9 bar, which is shallower than values inferred in regions outside the poles. If such a depletion exists, it would suggest that the polar region may be relatively well mixed below the $\sim$5 bar level. 
However, it should be noted that a homogeneous solution of ammonia vapor is also allowed if the temperature profile is moist adiabatic.
Nevertheless, the result here highlights that ammonia distribution in the region between the ammonia and water clouds behaves differently over the poles than at lower latitudes, and the mechanism responsible for ammonia depletion at lower latitudes either does not operate over the poles or does not deplete it as deeply there.

Various theories have been proposed to explain the origin of desiccated ammonia. 
\cite{guillotStormsDepletionAmmonia2020} proposed a "mush balls" theory in which downdraft convection originating from an ammonia-enriched thunderstorm can efficiently dry the air of ammonia. A meridional cell theory was also proposed to describe the differences in the ammonia gradient between Jupiter's zones and belts \citep{fletcherJupitersTemperateBelt2021, duerEvidenceMultipleFerrelLike2021}.
A model study suggested that depletion may be caused by processes associated with geostrophic adjustment after convection \citep{liMoistConvectionHydrogen2015}. \cite{liMoistAdiabatsMultiple2018} used a thermodynamic model modulated by a simple ‘stretch parameter’ and successfully simulated the depletion of Jovian gaseous species so that the profiles could be coordinated with the Galileo-probed profiles, with the surface of the distorted material being deflected to a deeper layer.
Descent flows can also bring depleted ammonia over the ammonia cloud base to a deeper layer \citep{depaterPeeringJupitersClouds2016}.

In addition to dynamical processes that may produce a vertical gradient in ammonia abundance, microphysical cloud processes can also contribute to the removal of NH$_3$ vapor over a broad pressure range. Previous studies have shown that ammonia can be depleted through the formation of an NH$_4$SH cloud near $\sim$2--3 bar, where NH$_3$ reacts with H$_2$S \citep{Briggs1989-md}, as well as through dissolution into aqueous solution droplets associated with the water cloud at deeper levels around $\sim$5--7 bar \citep{De-Pater1989-gt}. These processes operate at different altitudes and can both act to reduce the local NH$_3$ vapor abundance. In this study, we approximate their combined effects as a gradual decrease in NH$_3$ vapor over the pressure range where these cloud processes are expected to occur. This simplification is motivated by the broad vertical sensitivity of the Juno MWR channels used in our retrievals. As illustrated by the weighting functions (Fig.~\ref{Ext_weigthing_funcs}b), the contribution functions of these channels extend over several bars, limiting the ability of the observations to resolve sharp vertical transitions in the ammonia profile. Previous studies using VLA and Juno MWR data have retrieved more delicate vertical structures of ammonia depletion using flexible parameterizations, including stochastic-process-based approaches that typically require regularization constraints to stabilize the inversion \citep[e.g.,][]{liDistributionAmmoniaJupiter2017, moeckelAmmoniaAbundanceDerived2023}. The resulting fine-scale structures may not always be uniquely constrained given the broad vertical sensitivity of the observations. Under these conditions, a smooth depletion parameterization captures the combined effect of multiple NH$_3$ removal mechanisms without introducing additional poorly constrained parameters. Consequently, while detailed cloud microphysics may produce discrete depletion layers, their radiometric signatures would be difficult to distinguish from a gradual NH$_3$ gradient within the vertical resolution of the MWR observations.

\subsection{Deep water abundance}
The moist adiabat assumption suggests a deep water abundance of $1.8_{-1.1}^{+1.5}\times10^3$ ppmv, with a skewed uncertainty towards the larger values. The lifting condensation level is around 5 bar, which is in good alignment with the presence of an opaque cloud (blue shaded region in Fig~\ref{fig:nh3_comparison}b) observed in 5-$\mu$m spectroscopy at the GRS \citep{bjorakerGasCompositionDeep2018} and the North Equatorial Belt (NEB) \citep{bjorakerSpatialVariationWater2022}. Our retrieved deep water abundance falls below but within the 1$\sigma$ range of $2.5_{-1.6}^{+2.2} \times 10^3$ ppmv and $4.5\pm3.1 \times 10^3$ ppmv reported in the equatorial zone by \cite{liWaterAbundanceJupiters2020} and \cite{liSuperadiabaticTemperatureGradient2024}, respectively. Our estimate is greater than the updated value of 490$\pm$160 ppmv measured by the Galileo Probe Mass Spectrometer in the 5-$\mu$m hot spot \citep{wongUpdatedGalileoProbe2004}, which is a relatively dry region. These measurements demonstrate a varying but comparable deep water abundance across the planet.

Using Keck II/NIRSPEC 5-$\mu$m spectroscopy, \cite{bjorakerSpatialVariationWater2022} reported a water mole fraction of 1–-15 ppmv at the proposed water cloud top (4–-7 bar) in the NEB. This result is comparable to that derived from JIRAM measurements \citep{grassiSpatialDistributionMinor2020}, but is substantially lower than the deep water abundances retrieved in this study and in \cite{liWaterAbundanceJupiters2020} at a shallower layer. Although these near-infrared measurements cannot directly constrain the deep water abundance due to cloud attenuation, the deep cloud tops determined by these spectroscopic data imply a lower limit for the sub-cloud water abundance of approximately 0.5$\times$solar O/H ($\sim$420 ppmv) in the NEB \citep{wongDeepCloudsJupiter2023} and $\sim$1.1× solar ($\sim$930 ppmv) in the GRS \citep{bjorakerGasCompositionDeep2018}. These values are much closer to the MWR-derived abundances in both the north polar region and the equatorial zone. 

% Even in the absence of clouds, 5-$\mu$m measurements \citep{bjorakerSpatialVariationWater2022, grassiSpatialDistributionMinor2020, giles_cloud_2015} generally retrieved values much less than those detected by the GPMS and the Juno MWR, thus hinting at a significant steep vertical gradient in the water vapor concentration that extends well below 8 to 10 bar \citep{bjorakerSpatialVariationWater2022}. Such a gradient may also be associated with dynamical and microphysical processes, including the above-mentioned mushballs and Ferrell-like circulation theories. However, this possibility cannot be robustly assessed in the present study because the MWR channels exhibit considerably weaker sensitivity to water vapor than to ammonia.

\subsection{Cloud effect}
In this study, we do not include opacity from water cloud condensates in the radiative transfer calculations, considering the effect of cloud condensates is expected to be minimal at MWR frequencies.

Previous studies often adopt equilibrium cloud condensation models (ECCM) \citep{weidenschillingAtmosphericCloudStructures1973, Hueso2020ConvectiveStorms}. By construction, \cite{weidenschillingAtmosphericCloudStructures1973} (referred to hereinafter as WL73) adopts an idealized three-layer (H$_2$O, NH$_3$, NH$_4$SH) cloud structure that neglects precipitation processes, a primary control of cloud concentration in planetary atmospheres. As a result, cloud concentration predicted by WL73 ($\sim$10$^{-2}$--1 kg m$^{-3}$) is about four orders of magnitude greater than that predicted by self-consistent convection models which treat microphysical processes explicitly. For example, both \cite{Ge2024HeatFluxLimited} and \cite{SUGIYAMA2014Numericalsimulations} reported a cloud density that is on the order of $\sim$10$^{-6}$ kg m$^{-3}$. 

Based on the simulation and theory of \cite{Ge2024HeatFluxLimited}, the upper limit of the globally averaged cloud column density is independent of the atmospheric composition but is determined by the planetary heat flux and cloud lifetime. \cite{Wong2015FreshClouds} proposed a modification to the ECCM that introduces an updraft length scale needed to mathematically define the cloud density. The equations in WL73 explicitly set an updraft length scale of one pressure scale height, which is an exceptionally strong updraft. This is the reason why WL73 cloud densities are now considered excessive. Future implementations of the WL73 scheme should be applied with caution, as it likely represents an upper-limit, supply-driven regime associated with localized, transient storms, and may substantially bias the quantitative interpretation of observed spectra \citep{moeckelAmmoniaAbundanceDerived2023}. From that we estimate that the cloud water concentration on Jupiter is on the order of $\sim$10$^{-6}$ kg m$^{-3}$, which can be neglected by MWR observations.

\subsection{Temperature-ammonia degeneracy} The findings presented here indicate two potential solutions for the mean atmospheric state at Jupiter's north pole: (1) a dry adiabatic atmospheric temperature profile with ammonia depleted at a few bars, or (2) a vertically uniform ammonia abundance with a moist adiabatic temperature profile, allowing local warming at a few bars relative to the dry adiabat. Both of them reproduce the observations with comparable accuracy. These two scenarios differ at altitudes of several bars, with a temperature discrepancy of 2 K to 3 K, which is less than the lowest discernible range of absolute calibration uncertainty of 3.25 K ($\sigma(T_b)=1\%$), but fits well within the measurement uncertainty of $\sigma(\text{R(45$^\circ$)})=0.1\%$ (0.5 K random noise). Consequently, we cannot disentangle these two scenarios until (1) the temperature measurement around the 1-bar level is directly obtained by in situ measurements, (2) future radio occultation measurements could provide a constraint on the polar temperature, or (3) a joint analysis of JIRAM 5-$\mu$m observations could accurately constrain the ammonia concentration below the ammonia cloud deck.

It is worth noting that the two thermodynamic end-members adopted in this study represent only two simplified extreme scenarios and do not encompass the full range of physically plausible atmospheric states. Rather, they are adopted as illustrative cases to examine the sensitivity of the retrievals to different thermodynamic assumptions while keeping the parameter space consistent with the limited observational constraints. In this study, only five MWR frequencies are used, which effectively limits the number of independent atmospheric parameters that can be robustly constrained. Within this framework, the atmospheric structure can be reasonably characterized by a small number of parameters, such as the deep ammonia abundance, an ammonia depletion gradient, and a thermodynamic profile controlled by the assumed deep water abundance in the moist case. Although other atmospheric configurations are possible, for example, intermediate states between these two scenarios or profiles involving locally super-adiabatic layers (black-dashed line in Fig~\ref{fig:temperature_comparison}b), such configurations typically require additional degrees of freedom that cannot be uniquely constrained by the available observations, thus leading to increased degeneracy and potential overfitting.

\subsection{A warmer pole resembling Earth's tropics}
\begin{figure*}[htbp]
    \centering
    \includegraphics[width=0.9\linewidth]{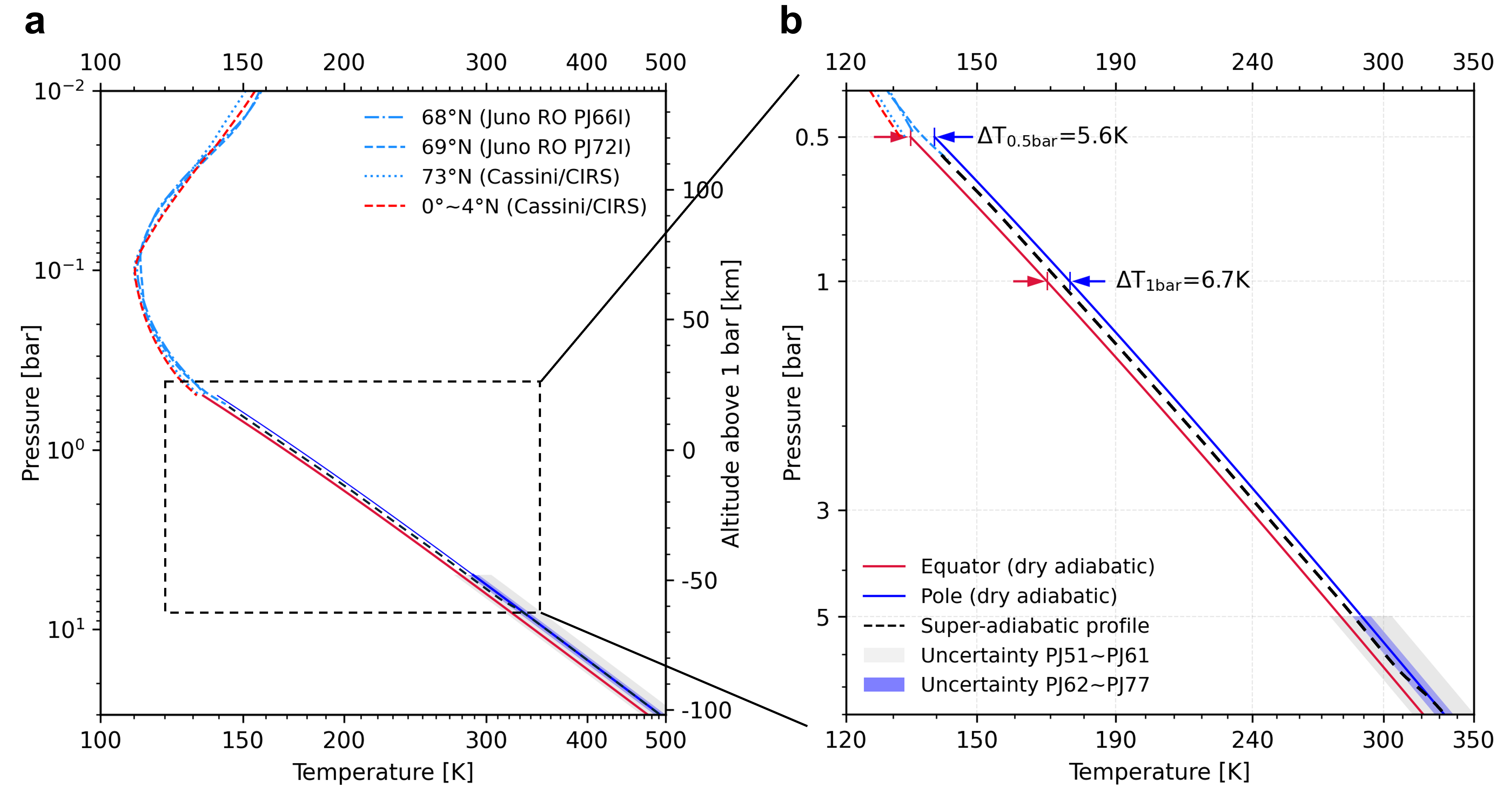}
    \caption{\textbf{Atmospheric thermal structure at Jupiter's north pole.} \textbf{a,} Temperature profiles at lower troposphere (0.5 to 30 bar) are presented for dry adiabatic retrieval at north pole (this study) and the Equator\citep{liWaterAbundanceJupiters2020}. Temperature profiles at upper troposphere and stratosphere are presented for Juno Radio Occultation (RO) measurements \citep{smirnovaProbingJupitersAtmosphere2025}, and for Cassini/CIRS data \citep{flasarIntenseStratosphericJet2004, simonmillerJupitersAtmosphericTemperatures2006}. \textbf{b,} Zoom-in plot of panel \textbf{a} for the pressure levels ranging from 8 bar to 0.4 bar. The temperature difference between the north pole and the Equatorial Zone at pressure levels of 0.5 and 1 bar are labeled. The gray and blue shades represent the 2$\sigma$ temperature uncertainty over the north pole during PJ51 to PJ61 and PJ62 to PJ77, corresponding to the observational random noise levels of 2.3 and 0.5 K, respectively. 
    The black dashed profile is shown as an illustrative example demonstrating how a modest deviation from the dry adiabatic structure (a super-adiabatic thermal structure) could reconcile the RO and MWR temperature profiles in the overlapping pressure range.}
    \label{fig:temperature_comparison}
\end{figure*}

Fig.~\ref{fig:temperature_comparison} shows that the temperature in the north pole is inferred to be approximately 5.6 K (at 0.5 bar) and 6.7 K (at 1 bar) higher than that assessed in the Equatorial Zone \citep{liWaterAbundanceJupiters2020}. We compared the dry adiabatic temperature profile because it provides a more direct measure of the deep interior potential temperature when surveying the global internal heat variability. This result implies either internal inhomogeneity within Jupiter or an enhanced polar heat flux relative to lower latitudes, as predicted by theoretical models \citep{jonathanConvectiveHeatTransfer2008,ingersollSolarHeatingInternal1978}. This finding also aligns with the more frequent occurrences of lightning activity in the north polar region \citep{brownPrevalentLightningSferics2018}, which is indicative of a preferential outward heat flux compared to lower latitudes.

If we are optimistic about the precision of the limb darkening measurement ($\sim$0.5 K), the resulting uncertainty in potential temperature is $\pm$3.25 K (Table~\ref{tab:tabS1}), which is precise enough to ascertain the higher temperature at the north pole with respect to the EZ at a two-sigma confidence level. However, in the present stage (PJ51-PJ61), there are still auroral features that might degrade the limb darkening measurement. Quantifying the effects is still ongoing work (Oyafuso et al., in prep). So, we show in Table~\ref{tab:tabS1} that, with a conservative noise level of 2.3 K and a systematic bias of 2\% quantified over the poles, the potential temperature is determined with an uncertainty of $\pm$6.2 K, which is still on the margin of confirming a $\sim$7 K higher temperature than the equatorial zone (1$\sigma$ confidence level). The precision in limb-darkening measurements is expected to improve over time as the observational geometry becomes more favorable and the sample size of auroral-free regions increases.

EZ is known to be locally one of the coldest regions in the upper troposphere and lower stratosphere, as evidenced by an increasing trend in 1-bar level temperature with latitude found from the combined analysis of Voyager radio occultation and Galileo Probe data \citep{guptaJupitersTemperatureStructure2022}. This warming trend across latitudes is also presented in Cassini/CIRS data reaching $\pm$73$^\circ$  (Fig.~\ref{fig:temperature_comparison}) and the Voyager/IRIS data reaching $\pm$55$^\circ$
\citep{flasarIntenseStratosphericJet2004, simonmillerJupitersAtmosphericTemperatures2006}, which is consistent with mid-infrared sensing of north polar temperatures in the same pressure regime observed by mid-infrared imaging \citep{bardetInvestigatingThermalContrasts2024}. Although these probings stop at 0.7 bar, our results are consistent with the possibility that the warming trend extends to deeper levels, at least to the 1-bar level. 

The latest Juno radio occultation (RO) measurements \citep{smirnovaProbingJupitersAtmosphere2025} reveal a warmer trend extending down to $\sim$0.56 bar in the upper troposphere near the pole (Fig.~\ref{fig:temperature_comparison}). These measurements are approximately 2 K to 3 K colder than our estimates over the north pole. Although we inferred a warmer deep temperature than EZ, temperatures above the water condensation level were constrained with larger uncertainty due to the temperature-ammonia degeneracy. Therefore, we show that an alternative super-adiabatic profile (Fig.~\ref{fig:temperature_comparison}b, black dashed) may be applied here to reconcile our results with those of the RO's at the upper troposphere. Such super-adiabaticity has previously been identified in Jupiter’s Equatorial Zone, where it represents a form of stable stratification due to water condensation in Jupiter's hydrogen-helium atmosphere \citep{liSuperadiabaticTemperatureGradient2024}. While a super-adiabatic temperature gradient may exist in the polar atmosphere, characterizing it would require additional parameters; therefore, it is not included as a retrieval end-member scenario in this study to avoid overfitting and parameter degeneracy.

It is intriguing to find that Jupiter’s north pole exhibits a thermal condition similar to Earth’s equator. On Earth, the troposphere is warmer at the tropics and colder at high latitudes, whereas in the stratosphere the tropics are colder. In addition, the stratosphere of Earth exhibits a monotonic increase in temperature from the tropics to the poles, while Jupiter shows oscillatory meridional variations, probably linked to its banded structure. What drives the colder equator on Jupiter? One possibility is the meridional circulation \citep{duerEvidenceMultipleFerrelLike2021}. The other possibility is internal heat flux, which is possibly linked to meridional circulation. We know that solar radiation does not penetrate deeper than $\sim$ 5 bar on Jupiter, so any influence below that level likely involves internal heat. However, the meridional distribution of internal heat is still unclear, despite earlier efforts \citep{ingersollSolarHeatingInternal1978, pirragliaMeridionalEnergyBalance1984}.

The fact that the pole is tied to a higher potential temperature at depths has wide implications, not only for its meteorology but also globally for interior science. To reconcile both gravity and spectroscopic constraints, the interior models posit a deep entropy associated with a potential temperature of approximately 10 K in excess of the values obtained by the Galileo probe \citep{miguelJupitersInhomogeneousEnvelope2022, howardExploringHypothesisInverted2023}. It is therefore tempting to assume that the potential temperatures measured at the poles are representative of those for the planet as a whole. One possibility could be that lower temperatures in nonpolar regions result from an inhibition of convection away from the planet’s rotation axis \citep{stevensonTurbulentThermalConvection1979}. If entropy (potential temperature) becomes independent of latitude at the hydrogen-helium phase separation region near 1 Mbar, the deep atmospheric temperatures as a function of latitude may be reconciled if nonpolar regions have an excess super-adiabaticity of order $4\times10^{-3}$. The implications for the global interior structure and the planet’s evolution will have to be studied.

\begin{acknowledgments}
We sincerely thank the editor and reviewers for their constructive comments and feedback on the manuscript.
J.H., C.L., and S.K.A. are supported by the NASA Juno Program, under NASA Contract NNM06AA75C from the Marshall Space Flight Center, through subcontract 699056KC and Q99063JAR to the University of Michigan from the Southwest Research Institute.
Some of this research was carried out at the Jet Propulsion Laboratory, California Institute of Technology, under a contract with the National Aeronautics and Space Administration (80NM0018D0004). 
L.N.F. is supported by the Science and Technology Facilities Council Consolidated Grant reference ST/W00089X/1.
\end{acknowledgments}

\begin{contribution}
C.L. and J.H. conceptualized the work and designed the methodology. J.H. carried out the retrievals and conducted the analysis. C.L. developed the cloud equilibrium condensation model and the microwave radiative transfer module. Z.Z. and F.A.O. processed the Juno MWR raw data and advised on the uncertainty assessment. Y.K., E.G. and M.S. provided the Juno radio occultation data. S.K.A., L.N.F., T.G., Y.K., L.L., Y.L., A.M., G.S.O., J.H.W and M.H.W. participated in manuscript discussions and revisions. C.L., S.M.L. and S.J.B. supervised the work within the Juno mission atmosphere working group. J.H. wrote the manuscript with input from all authors.
\end{contribution}

\section*{Code and Data Availability}
Juno MWR data are archived in the NASA Planetary Data System on the Planetary Atmospheres Node at \url{https://pds-atmospheres.nmsu.edu/data_and_services/atmospheres_data/JUNO/microwave.html}. 
The code associated with data reduction and retrieval can be accessed at the GitHub repository \url{https://github.com/jihenghu/Jupiter_northpole}. The code version 1.0 is publicly available at \url{https://doi.org/10.5281/zenodo.19056004}, together with the table of angular dependence coefficients of the polar-mean brightness temperatures.

\software{Canoe (https://github.com/chengcli/canoe),  
          SPICE (https://naif.jpl.nasa.gov/naif/toolkit.html), 
          emcee (https://emcee.readthedocs.io),
          }
\facility{Juno (MWR)} 

\clearpage

\appendix

\renewcommand{\thefigure}{\Alph{section}\arabic{figure}} % Reset numbering to A1, A2, etc.
\renewcommand{\thetable}{\Alph{section}\arabic{table}} % Reset numbering to A1, A2, etc.

\section{Bayesian statistical results}\label{appendix_MCMC}
\setcounter{figure}{0}

\begin{figure}[ht]
    \centering
    \includegraphics[width=0.85\linewidth]{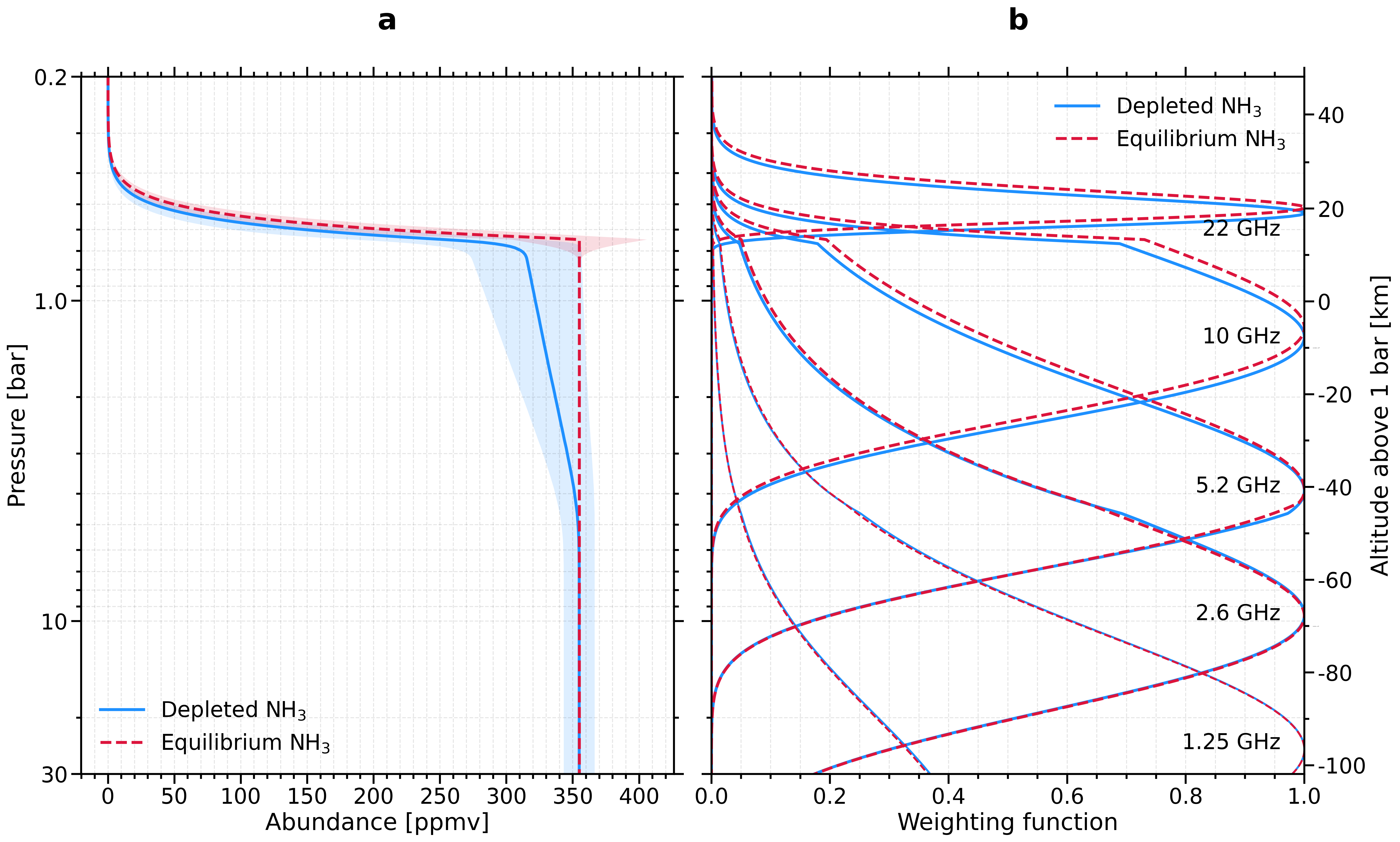}
    \caption{\textbf{Ammonia profiles and the corresponding atmospheric weighting functions}. \textbf{a,} Ammonia profiles of the equilibrium (dashed red) and depletion (solid blue) models retrieved in the main-text, with the corresponding standard deviation envelopes denoted by the red and blue shaded regions, respectively. \textbf{b,} Atmospheric weighting functions of equilibrium (dashed red) and depletion (solid blue) ammonia models at six MWR frequencies.}
    \label{Ext_weigthing_funcs}
\end{figure}

\begin{figure}[ht]
    \centering
    \includegraphics[width=0.85\linewidth]{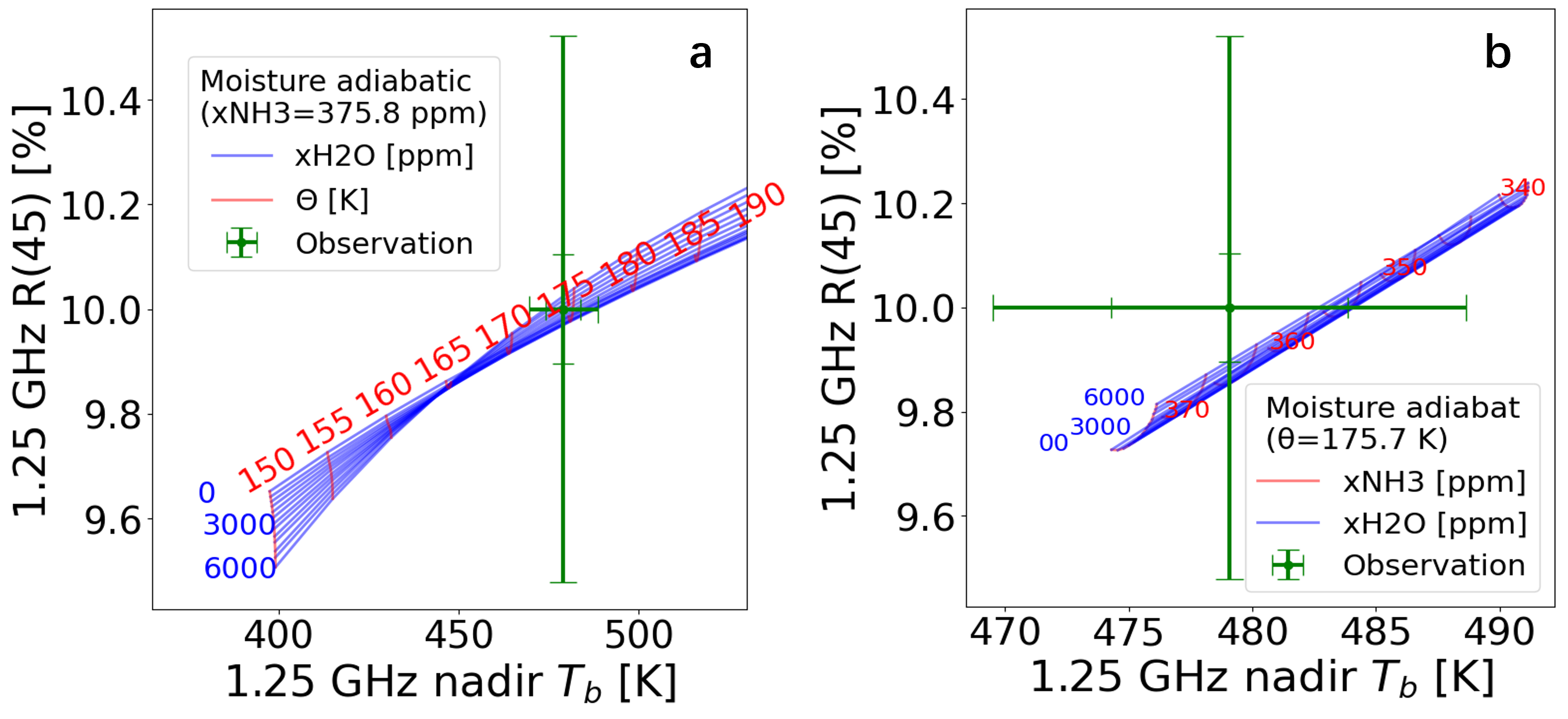}
    \caption{\textbf{Sensitivity tests of the 1.25 GHz nadir brightness temperature and limb darkening to variations in deep atmospheric parameters in the moist-adiabat model.} Panels\textbf{ a} and \textbf{b} show the dependence of the nadir brightness temperature and the limb-darkening ratio (R45) on the deep potential temperature and deep water and ammonia abundance. The green error bars represent the observational uncertainty levels, corresponding to 1\% and 2\% of the nadir brightness temperature, and 0.5 and 2.3 K for the limb-darkening ratio. The blue contours indicate levels of deep water abundance in 500-ppmv increments. The red contours denote potential temperature in 5-K increments in panel \textbf{a}, and deep ammonia abundance in 5-ppmv increments in panel \textbf{b}.}
    \label{fig:ext_sensi_xH2O}
\end{figure}

\begin{figure}[ht]
    \centering
    \includegraphics[width=0.8\textwidth]{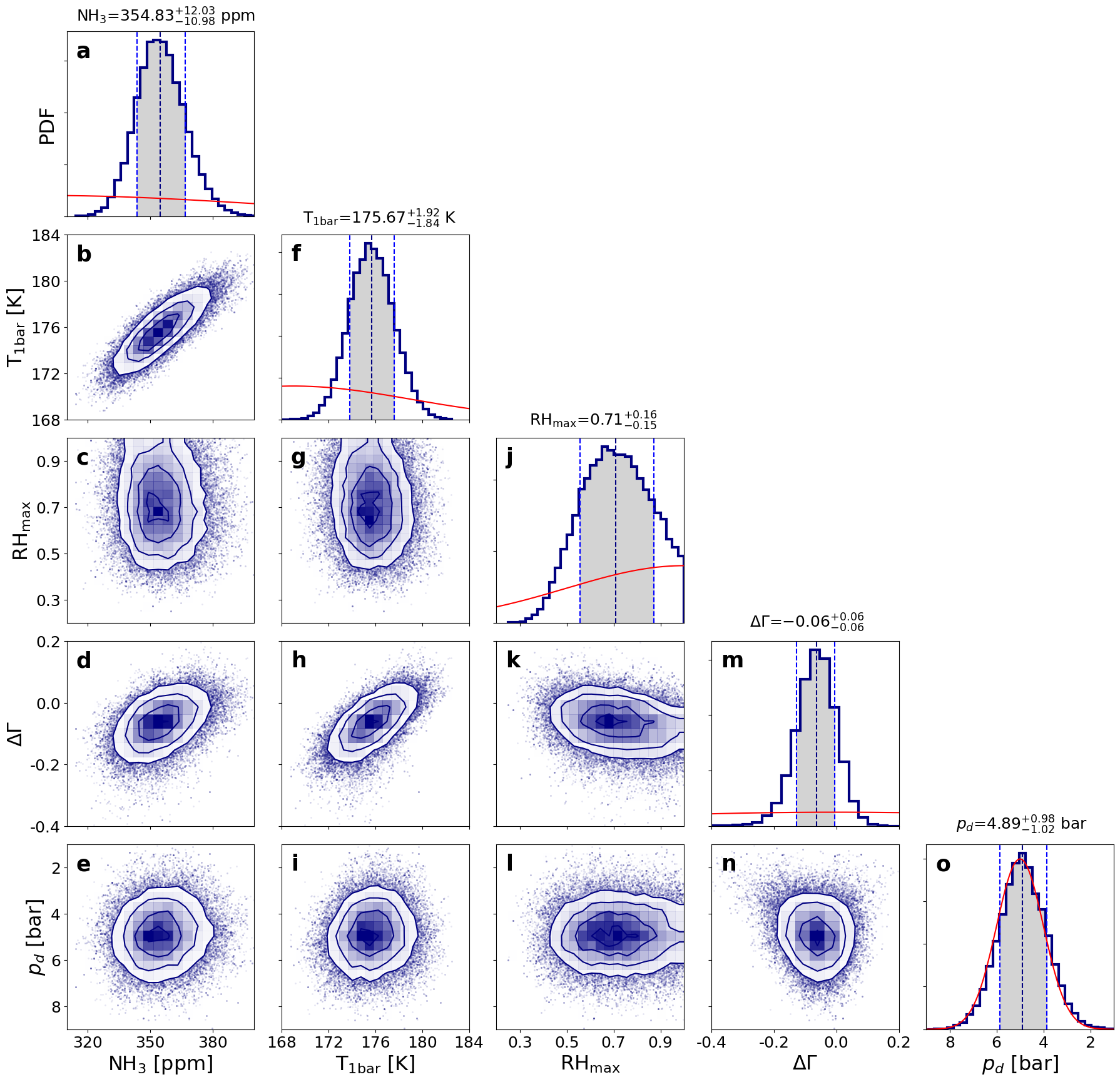} 
    \caption{\textbf{Corner plot summary of the posterior probability distribution retrieved under the dry adiabatic assumption in the main text.} The scatter denotes each of the last 7000 states Markov chain, contours illustrate the joint probability distributions between each two of the inferred parameters. The marginal probability distribution for each parameter is presented with the one-dimensional histogram at the top plot of each column, where the red line denotes the prior distribution and the navy step-shaped line is the posterior distribution. The blue dashed lines represent the 16th, 50th and 84th percentiles of the samples (1$\sigma$ uncertainty) in the marginal distribution.}
    \label{Ext1_MCMC_dry_corners}
\end{figure}

\begin{figure}[ht]
    \centering
    \includegraphics[width=0.5\textwidth]{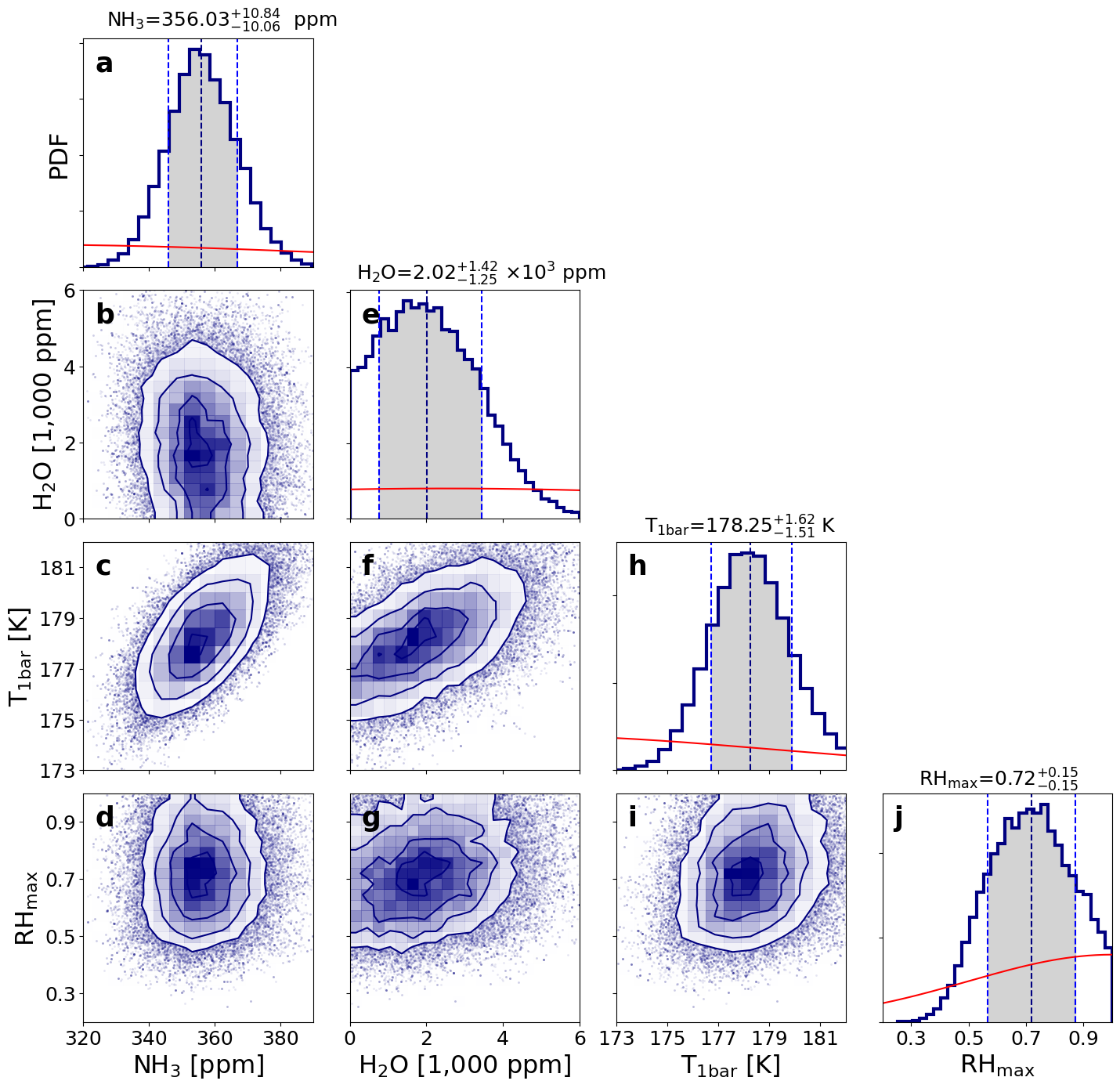} 
    \caption{\textbf{The same to Fig.~\ref{Ext1_MCMC_dry_corners}, but for MCMC inversion under the moist adiabatic assumption.} 
    }
    \label{Ext2_MCMC_moist_corners}
\end{figure}

\begin{figure}[ht]
    \centering
    \includegraphics[width=0.8\textwidth]{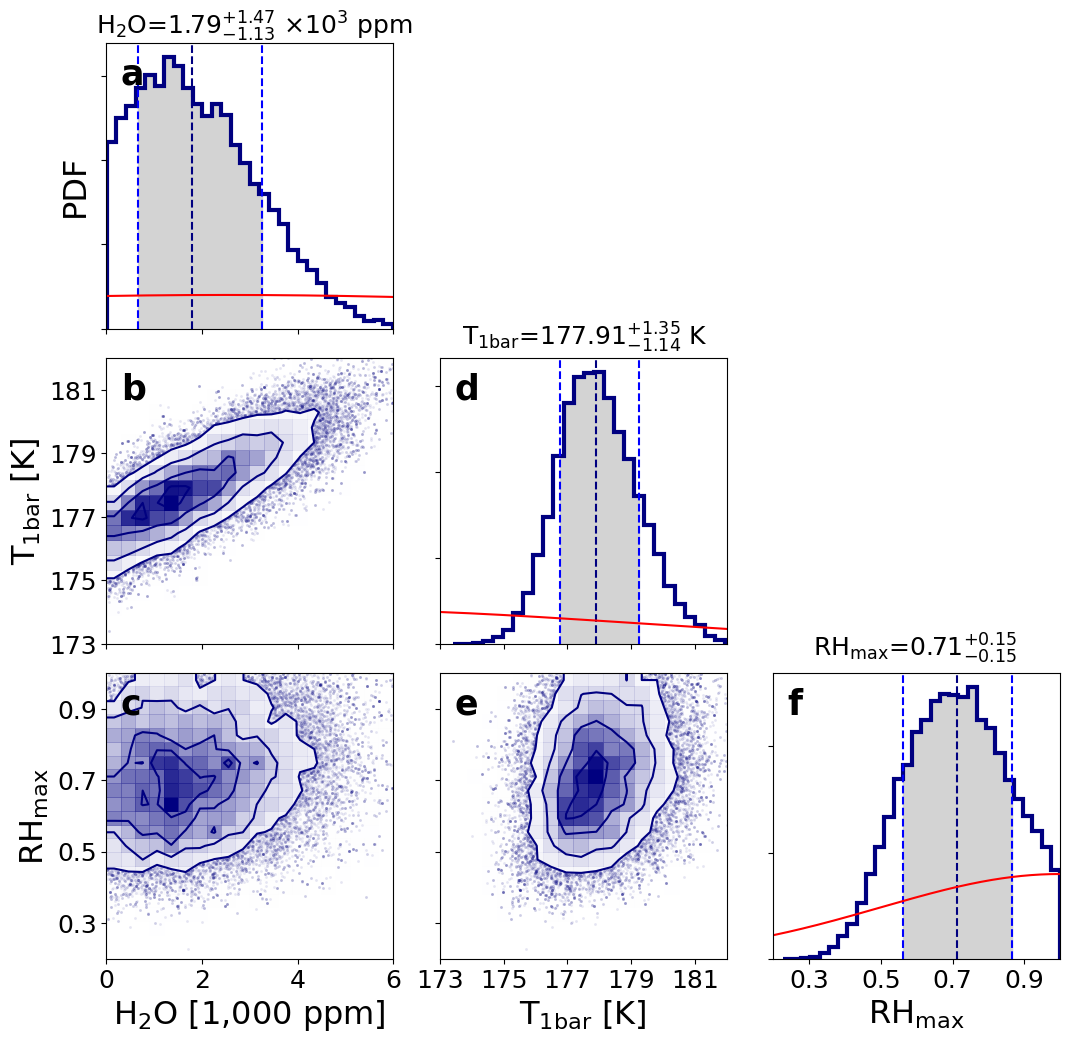} 
    \caption{\textbf{Corner plot of MCMC test retrieving deep ammonia abundance together with all parameters in moist adiabatic assumption.}
    The corresponding deep potential temperature is $\theta=176.1_{-1.5}^{+1.6}$ K. By varying deep ammonia abundance in the moist-adiabat retrieval, MCMC results in estimates of all parameters closely consistent with those inferred by fixing for moist-adiabat case the deep ammonia abundance obtained from the dry-adiabat case (Figure \ref{Ext2_MCMC_moist_corners}). The retrieved deep ammonia abundance is $356.0_{-10.1}^{+10.8}$ ppm compared to $354.8_{-11.0}^{+12.0}$ ppm for the dry-adiabat case and the deep potential temperature is $176.1_{-1.5}^{+1.6}$ compared to $175.7_{-1.8}^{+1.9}$ for the dry-adiabat case (Figure \ref{Ext1_MCMC_dry_corners}). This demonstrates that MWR 1.25 GHz channel observations dictate the deep atmospheric properties, and the higher frequency channels govern the shallow variations, which can either be ammonia or temperature.
    }
    \label{sensitive_moist_NH3}
\end{figure}

\begin{figure}[ht]
    \centering
    \includegraphics[width=0.7\linewidth]{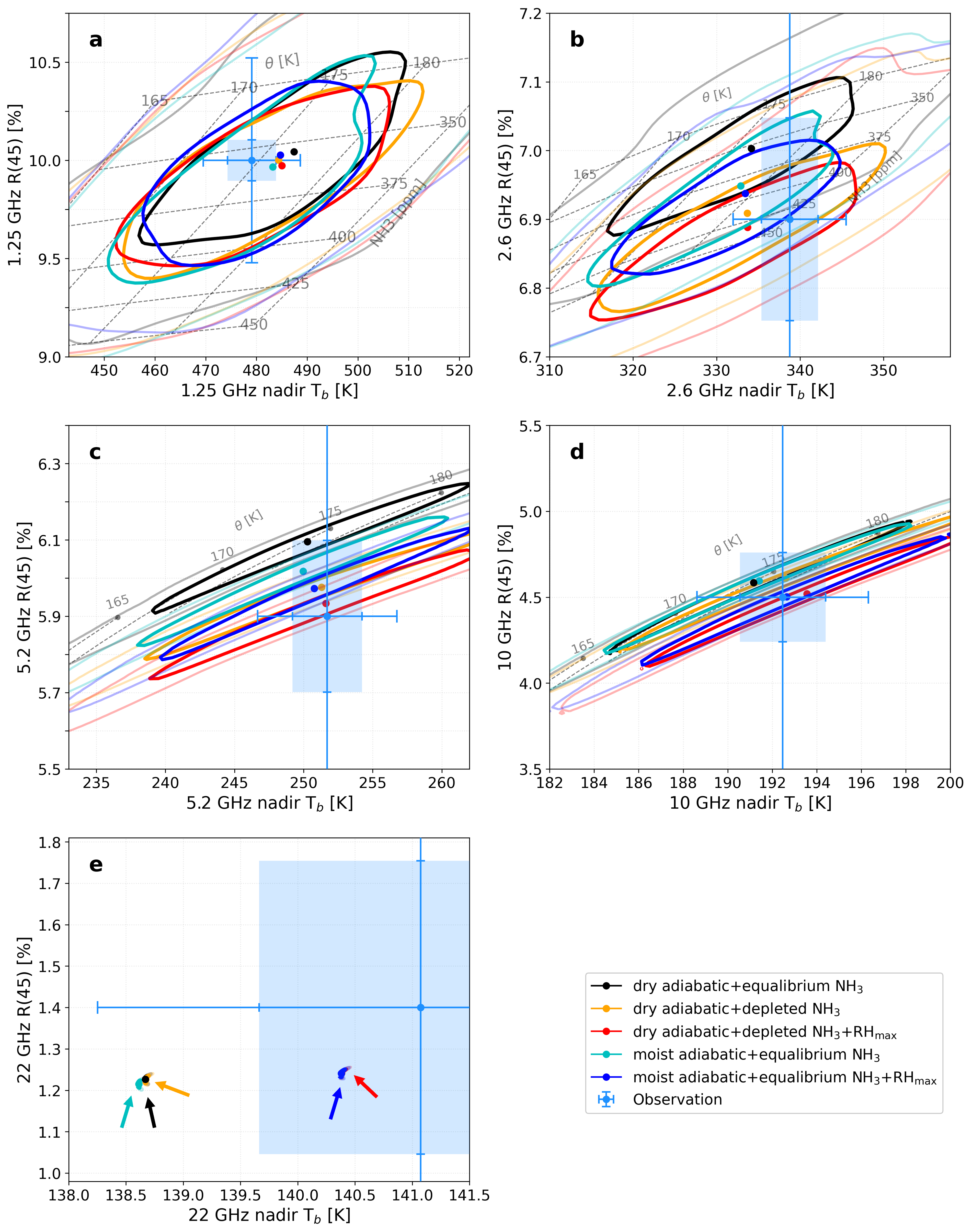}
    \caption{\textbf{Fitting performance of different atmospheric structure assumptions over Jovian north pole.} Grey dashed lines in \textbf{a} and \textbf{b} represent the simulated nadir $T_b$ and $R45$ as functions of deep-layer ammonia abundance, $x\text{NH}_3$, and potential temperature, $\theta$, under the dry adiabatic assumption. Light blue markers denote the MWR observations for each channel, with error bars representing the measurement error intervals of 1\% and 2\% in nadir $T_b$, 0.5 and 2.3 K in $R45$. The joint probability distributions of different structure assumptions were derived from 1000-step Monte Carlo sampling, where forward simulations were performed with perturbations of 1\% for $T_b$ and 0.5 K for $R45$. For each probability distribution, the fully opaque line denotes the 1$\sigma$ uncertainty interval and the partially opaque line denotes the 2$\sigma$ uncertainty interval. In panel \textbf{e}, colored arrows are used to indicate the locations of different atmospheric assumptions. At 22 GHz, Tb is minimally sensitive to deep-layer ammonia abundance and potential temperature. As a result, both the grey dashed model curves and the Monte Carlo sampled probability distribution curves collapse into a narrow region around the plotted points. 0.6-GHz channel is not shown as it is not involved in inversion. }
    \label{Ext3_LUT_dry_adb}
\end{figure}

\clearpage
\section{Formal Uncertainty analysis}\label{appendix_uncertainty}
\setcounter{figure}{0}
\setcounter{table}{0}

The uncertainty of the retrieved potential temperature largely depends on our understanding of the uncertainty in limb darkening induced by measurement noise. Though the nominal instrument noise level of MWR is 0.5 K, the limb darkening estimated in the polar region is expected to suffer from a greater magnitude of uncertainty than nominal, considering the auroral impacts in most orbits. 

A direct estimation of the standard deviation of the limb darkening at 45° (LD45) using all samples from eleven orbits (Table~\ref{tab:perijovetable}) yields an unrealistically low value, on the order of 0.03 K. Instead, we assess the dispersion in limb darkening over orbits PJ51 to PJ54 to derive a more realistic estimate of the noise level.

\begin{table}[htbp]
\centering
\caption{\textbf{Mission Perijove dates, latitudes, and longitudes used in this paper.} Source: https://www.missionjuno.swri.edu/mission-perijoves}
\label{tab:perijovetable}
\begin{tabular}{cccc}
\hline
\textbf{Perijove} & \textbf{Date} & \textbf{Latitude (Planetocentric)} & \textbf{Longitude* (System III West)} \\
\hline
51 & 2023 May 16 & 43\textdegree & 140\textdegree \\%\hline
52 & 2023 Jun 23 & 44\textdegree & 80\textdegree \\%\hline
53 & 2023 Jul 31 & 45\textdegree & 120\textdegree \\%\hline
54 & 2023 Sep 7  & 45\textdegree & 190\textdegree \\%\hline
55 & 2023 Oct 15 & 46\textdegree & 110\textdegree \\%\hline
56 & 2023 Nov 22 & 47\textdegree & 120\textdegree \\%\hline
57 & 2023 Dec 30 & 47\textdegree & 90\textdegree \\%\hline
58 & 2024 Feb 3  & 48\textdegree & 290\textdegree \\%\hline
59 & 2024 Mar 7  & 49\textdegree & 0\textdegree \\%\hline
60 & 2024 Apr 9  & 50\textdegree & 40\textdegree \\%\hline
61 & 2024 May 12 & 51\textdegree & 250\textdegree \\
\hline
\end{tabular}

\vspace{0.5em}
\raggedright\textsuperscript{*} Longitudes rounded to the nearest 10\textdegree.
\end{table}

\begin{figure}[ht]
    \centering
    \includegraphics[width=\linewidth]{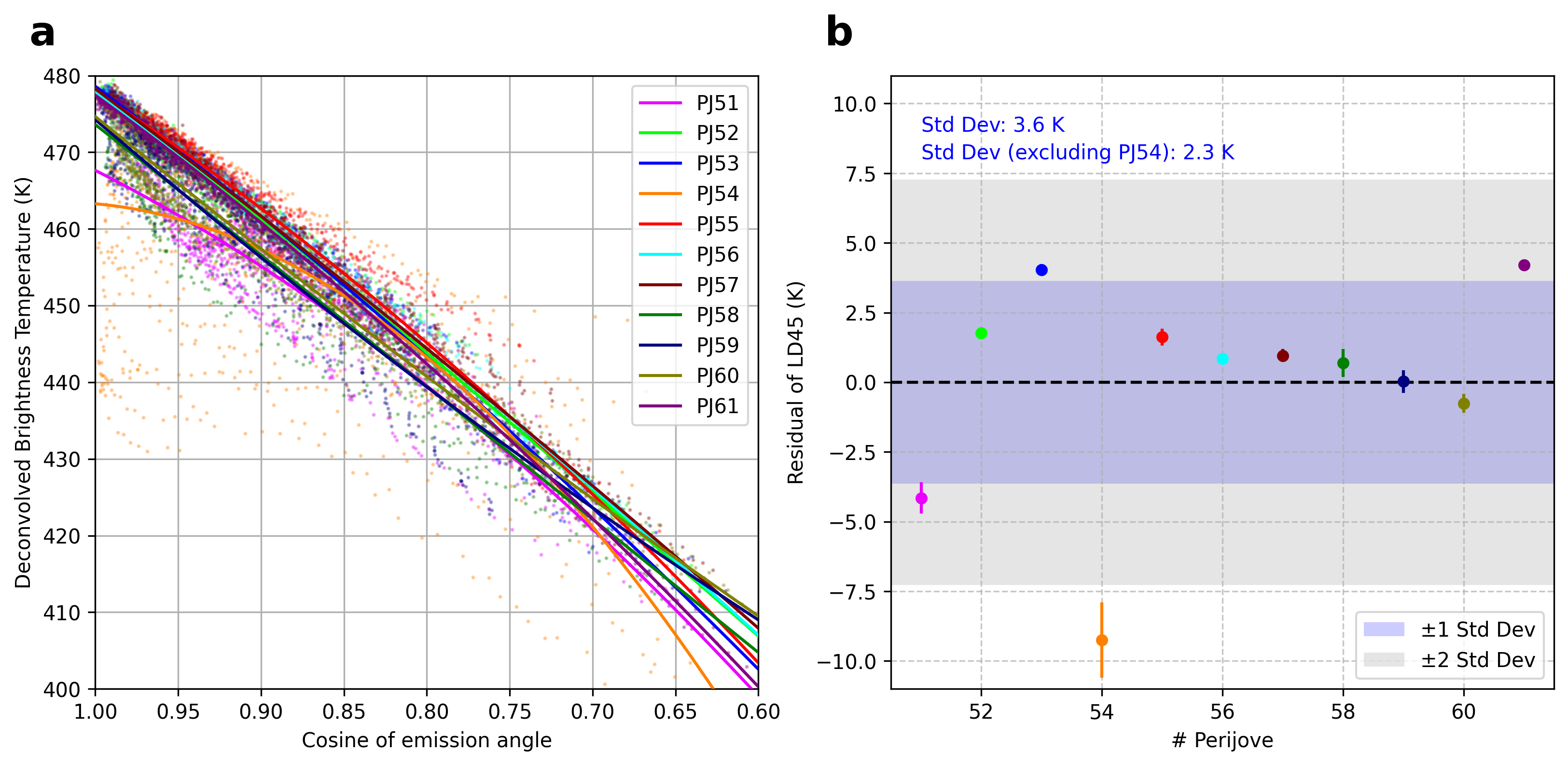}
    \caption{\textbf{Assessment of uncertainty of limb darkening estimated from the eleven Juno orbits.} Colored dots in panel \textbf{a} represent each sample of deconvolved brightness temperatures of different orbits. The lines are fits versus emission angle using the quadratic angular dependence model (Eq.\ref{eq:tb_model}). Dots in panel \textbf{b} represent the residual of limb darkening value derived from the fit of each orbit relative to the multiple-perijove mean value. Orbit-wise standard deviation of residual LD45 is computed including and excluding PJ54.}
    \label{fig:residual_ld45}
\end{figure}

\begin{figure}[ht]
    \centering
    \includegraphics[width=0.7\linewidth]{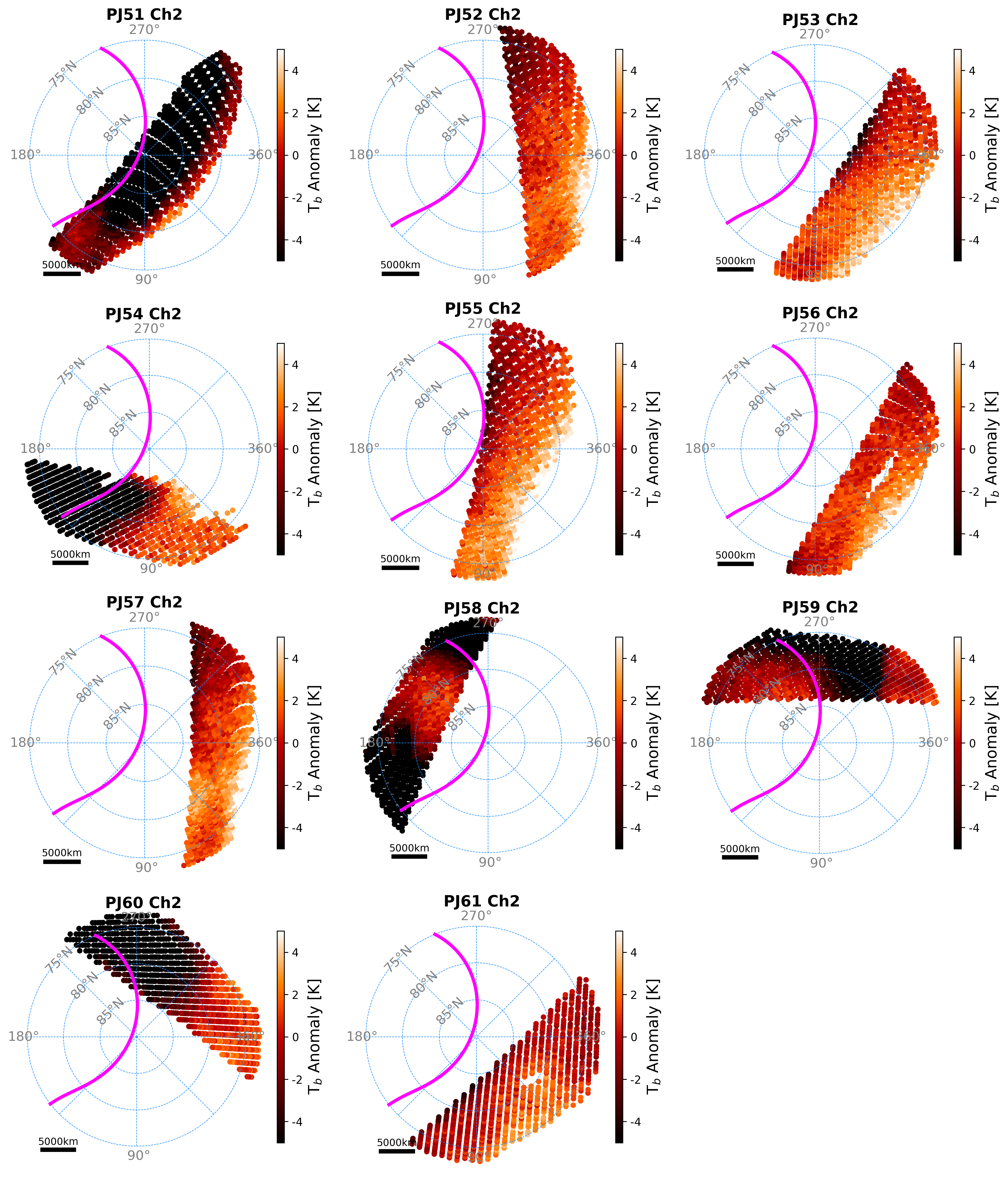}
    \caption{\textbf{Residual maps of 1.25-GHz brightness temperature of all eleven perijoves.} Magenta line represents the north part of main aurora oval, around which  brightness temperatures are significant suppressed due to the impact of auroral emission.}
    \label{Ext_aurora_perijove}
\end{figure}

Fig.~\ref{fig:residual_ld45}a displays the deconvolved brightness temperatures for individual observation points. For each orbit, we fit the data using the quadratic angular dependence model (Eq.\ref{eq:tb_model}) and then compute the LD45 values, which are shown for each perijove in Fig.~\ref{fig:residual_ld45}b. Orbits PJ51 and PJ54 exhibit notably lower LD45 values, consistent with the fact that large portions of these orbits lie near the main auroral oval (Fig.~\ref{Ext_aurora_perijove}). In contrast, the other orbits are less affected by aurora and exhibit moderate to high LD45 values.

The standard deviation of LD45 across these eleven orbits is 3.6 K, an unacceptably high estimate given that some orbits are strongly contaminated by auroral emissions.
To obtain a more representative estimate of the uncertainty, we exclude PJ54, whose LD45 falls outside the 2$\sigma$ range, and recompute the standard deviation. The revised value is 2.3 K (0.5 \%), which we adopt as a conservative estimate of the noise level throughout this study.

The radiative transfer simulations shown in main-text Fig.~\ref{fig-spectra}b reveal that, in channel of 1.25 GHz, both nadir brightness temperature ($T_b$, units K) and limb darkening at 45$^\circ$ (R45, units \%) respond linearly to both deep ammonia abundance ($x\text{NH}_3$, units ppmv) and the potential temperature ($\theta$, units K). Thus, the following matrix of partial derivatives (coefficients) can be determined using the multiple linear regression (MLR) method,

\begin{equation}
    \begin{bmatrix} 
\frac{\partial \theta}{\partial R45} & \frac{\partial \theta}{\partial \text{T}_b} \\ 
\frac{\partial x\text{NH}_3}{\partial R45} & \frac{\partial x\text{NH}_3}{\partial \text{T}_b} 
\end{bmatrix}  =
\begin{bmatrix} 
-10.73 & 0.32 \\ 
-77.54 &  0.30 
\end{bmatrix}. 
\end{equation}

According to the propagation of the uncertainty, the uncertainties of $x\text{NH}_3$ and $\theta$ introduced by typical uncertainty levels of R45 (\%) and $T_b$ ($\sim$ 480 K) can be determined, as shown in Table~\ref{tab:tabS1}. %is

\begin{table}[ht]
\centering
\caption{\textbf{Uncertainty analysis of deep ammonia ($x\text{NH}_3$) and potential temperature ($\theta$).}  Uncertainties are presented as functions of different accuracy levels of nadir brightness temperature ($T_b$) and limb darkening (R45) at 1.25 GHz.}
\label{tab:tabS1}%

\begin{tabular*}{\textwidth}{@{\extracolsep\fill}cccc} % Use @{} to reduce extra padding
\hline
 $\sigma$($\theta$) & $\sigma(T_b)$=1\% (4.8 K)  & $\sigma(T_b)$=2\% (9.6 K) \\ 
\hline
% \textbf{$\theta$ (K)}\\
 $\sigma(R45)$=0.5 K (0.1\%) & 1.87         & 3.25  \\ 
 $\sigma(R45)$=2.3 K (0.5\%)  & 5.56           & 6.17  \\ 
\hline

 $\sigma$($x\text{NH}_3$) &  &  \\ 
 \hline
 $\sigma(R45)$=0.5 K (0.1\%) & 7.8        & 8.3  \\ 
 $\sigma(R45)$=2.3 K (0.5\%)  & 38.8            & 38.9  \\ 
\hline
\end{tabular*}
\end{table}

In the optimal accuracy assurance case, where $\sigma(\text{R45})=0.1\%$ ($\sim$0.5 K random noise) and  $\sigma(\text{T}_b)=1\% $ ($\sim$4.8 K systematic bias), we get
\begin{equation}
\sigma(\theta) = \sqrt{-10.73^2 \cdot 0.1^2 + 0.32^2 \cdot (480 \cdot 1\%)^2 }\approx1.87\, \text{K},
\end{equation}
which means a difference larger than $\Delta\theta$=3.8 K is distinguishable with 95.45\% (2$\sigma$) confidence.

\bibliography{references,science_template}
\bibliographystyle{aasjournalv7}

\end{document}